\newif\ifrevtex
\IfFileExists{revtex4-2.cls}{\revtextrue}{\revtexfalse}
\ifrevtex
\def\paperclass{\documentclass[aps,prapplied,twocolumn,superscriptaddress,nofootinbib,longbibliography]{revtex4-2}}
\else
\def\paperclass{\documentclass[10pt,twocolumn]{article}}
\fi
\paperclass
\ifrevtex\else\usepackage[margin=1.8cm]{geometry}\fi
\usepackage{amsmath,amssymb,amsthm,mathtools,bm,graphicx,booktabs,xcolor}
\usepackage[caption=false]{subfig}   
\usepackage{hyperref}
\hypersetup{colorlinks=true,linkcolor=blue!50!black,citecolor=blue!50!black,urlcolor=blue!50!black}
\usepackage{orcidlink}
\usepackage{tikz}
\usetikzlibrary{positioning,arrows.meta,calc,fit,backgrounds}
\usepackage{quantikz}
\usepackage{algpseudocode}   
\newcounter{algorithm}

\newcommand{\algtitle}[1]{\refstepcounter{algorithm}\noindent\rule{\linewidth}{0.8pt}\par\noindent\textbf{Algorithm~\thealgorithm}\quad #1\par\vspace{-0.4\baselineskip}\noindent\rule{\linewidth}{0.4pt}\par}
\newcommand{\algend}{\par\vspace{-0.4\baselineskip}\noindent\rule{\linewidth}{0.4pt}}
\ifrevtex\else\newenvironment{acknowledgments}{\section*{Acknowledgments}}{}\fi

\newtheorem{theorem}{Theorem}
\newtheorem{proposition}{Proposition}
\newtheorem{corollary}{Corollary}
\newtheorem{definition}{Definition}
\newcommand{\nm}{n_{\mathrm{m}}}
\newcommand{\nw}{n_{\mathrm{w}}}

\newcommand{\MMD}{\mathrm{MMD}}
\newcommand{\Tr}{\operatorname{Tr}}
\newcommand{\Lam}{\Lambda}

\newcommand{\E}{\mathbb{E}}

\newcommand{\Id}{\mathbb{I}}

\providecommand{\bra}[1]{\langle #1|}
\providecommand{\ket}[1]{|#1\rangle}

\newlength{\figw}
\newcommand{\figfile}[2]{\IfFileExists{figures/#1}{\includegraphics[width=#2]{figures/#1}}{\setlength{\figw}{#2}\fbox{\parbox{0.86\figw}{\centering\footnotesize\textcolor{red}{figure not yet produced}\\\tiny\texttt{\detokenize{#1}}}}}}
\newsavebox{\fitbox}\newcommand{\fitcol}[1]{\sbox{\fitbox}{#1}\ifdim\wd\fitbox>\columnwidth\resizebox{\columnwidth}{!}{\usebox{\fitbox}}\else\usebox{\fitbox}\fi}

\newcommand{\UAUCfloor}{0.479}
\newcommand{\UBeatSentence}{every coherent run beats every dephased and every classical run on held-out data}
\newcommand{\UCeil}{91.5\%}
\newcommand{\UChiEquiv}{6.2}
\newcommand{\UCohMean}{0.786}
\newcommand{\UCohWorst}{0.788}
\newcommand{\UCommonSE}{0.09}
\newcommand{\UDiffComb}{-0.067\pm0.024}
\newcommand{\UDiffDeph}{0.104\pm0.009}
\newcommand{\UDiffHmm}{0.103\pm0.010}
\newcommand{\UDiffTreeFour}{-0.005\pm0.019}
\newcommand{\UDiffTreeSix}{+0.033\pm0.019}
\newcommand{\UDigitsPass}{3}
\newcommand{\UGainDeph}{13\%}
\newcommand{\UGainHmm}{13\%}
\newcommand{\ULambdaStar}{0.05}
\newcommand{\UMixAUC}{0.541}
\newcommand{\UMixIntended}{68\%}
\newcommand{\UN}{1\,797}
\newcommand{\UOtherBest}{0.864}
\newcommand{\USeeds}{3}
\newcommand{\Ud}{56}

\providecommand{\UN}{1\,797}\providecommand{\Ud}{56}

\begin{document}
	
	\title{Quantum Generative Modeling with a Coherent Memory, Exact Training and a Memory-Dimension Advantage}
	
	\ifrevtex
	\author{Jamal Slim\orcidlink{0000-0002-9418-8459}}
	\email{jamal.slim@desy.de}
	\affiliation{Deutsches Elektronen-Synchrotron DESY, 22603 Hamburg, Germany}
	
	\author{Saverio Monaco\orcidlink{0000-0001-8784-5011}}
	\affiliation{Deutsches Elektronen-Synchrotron DESY, 22603 Hamburg, Germany}
	\affiliation{RWTH Aachen University, 52062 Aachen, Germany}
	
	\author{Ran Xue\orcidlink{0000-0002-2009-6279}}
	\affiliation{Deutsches Elektronen-Synchrotron DESY, 22603 Hamburg, Germany}
	
	\author{Dirk Kr\"ucker\orcidlink{0000-0003-1610-8844}}
	\affiliation{Deutsches Elektronen-Synchrotron DESY, 22603 Hamburg, Germany}
	
	\author{Kerstin Borras\orcidlink{0000-0003-1111-249X}}
	\affiliation{Deutsches Elektronen-Synchrotron DESY, 22603 Hamburg, Germany}
	\affiliation{RWTH Aachen University, 52062 Aachen, Germany}
	\else
	\author{Jamal Slim, Saverio Monaco, Ran Xue, Dirk Kr\"ucker, Kerstin Borras\\ \small Deutsches Elektronen-Synchrotron DESY, Hamburg, and RWTH Aachen University}
	\date{Preview}
	\maketitle
	\fi
	\begin{abstract}
		We introduce the coherent memory Born machine (CoMB), an algorithm for quantum generative modeling on a register whose size does not depend on the dimension of the data. CoMB generates a sample one component per round. In each round, a work register is entangled with a memory register and measured, and the outcome gives one component. The memory is never measured and carries correlations between rounds. We show that the characteristic function of the generated distribution is a product of small matrices, one per round. This gives exact training, with maximum mean discrepancy losses and gradients costing time linear in the data dimension, without sampling or likelihood. We also compile every round onto one work qubit and evaluate device noise exactly. To isolate coherence, each CoMB is compared with its dephased twin and with equally trained classical hidden Markov models. We benchmark it on continuous data, $368$-cell calorimeter showers from five qubits, and on sequential data, handwritten digits from three qubits. The coherent memory provides a memory-dimension advantage. Four coherent levels do the work of about six classical states on both benchmarks, and of at least five with $95\%$ confidence on showers at every training budget, including convergence. On showers, removing their coherence raises the loss by $23$ to $31\%$. With eight levels, where all models approach a common floor, coherent and classical memories become indistinguishable with longer training. On an IBM Heron processor, the memory carries two thirds of the inter-round correlation through twelve mid-circuit measurements.
	\end{abstract}
	
	\ifrevtex\maketitle\fi
	
	\section{Introduction}\label{sec:intro}
	
	Generative models learn a probability distribution from examples and produce new samples from it. Quantum generative models are attractive for this task for two reasons. Born-rule sampling produces correlated outputs without an explicit sampling procedure~\cite{Benedetti2019,Liu2018,Coyle2020}, and quantum memories can represent some stochastic processes with less memory than any classical model~\cite{Gu2012,Monras2010,Elliott2020,Wu2023}. Whether either property yields a measurable advantage on realistic, high-dimensional data, under a training procedure that itself scales, is an open question. Its answer matters for every application of quantum generative modeling, from image and sequence generation to fast scientific simulation, such as the simulation of particle showers in collider detectors~\cite{geant4,atlashllhc,paganini,krause,calochallenge}.
	
	Three obstacles stand in the way. First, the quantum circuit Born machine~\cite{Benedetti2019,Liu2018,Coyle2020} holds an entire sample in its register, so a sample of $d$ values at $b$ bits each requires $bd$ entangled qubits. Applications to high-dimensional data therefore compress the data into a few latent variables or assemble samples from independently generated patches~\cite{qfan,DualPQC2021,CaloQVAE2024}, and correlations removed by the compression or spanning several patches cannot be generated. Second, these models are trained with losses estimated from samples of the model, which adds shot noise to every gradient and compounds the trainability problems of variational circuits~\cite{Liu2018,Rudolph2023}. The likelihood is no alternative on hardware, because a device returns samples rather than probabilities. Other approaches change the model class~\cite{Zhang2024QuDDPM} or train classically and deploy quantumly~\cite{Recio2025}. Third, claims of quantum advantage in generative modeling are rarely tested against controls that isolate the quantum resource. Whenever a loss constrains only low-order correlators, a classical surrogate can match the quantum model~\cite{BornUltimatum2025}, so an observed advantage must be attributed with care.
	
	Sequential models remove the first obstacle by construction. A memory system that interacts with one site at a time can prepare correlated states of arbitrary length~\cite{Schon2005,FossFeig2021}, and hidden quantum Markov models (HQMMs), which emit one classical symbol per interaction, can represent processes that no classical hidden Markov model of the same dimension can~\cite{Monras2010,Adhikary2020}. Their expressivity is understood through the positive semidefinite rank of their cut matrices~\cite{fawzi2015} and the quantum dimension needed to reproduce a classical process~\cite{QDimRed2026}. In tensor-network terms their distributions are locally purified matrix product states~\cite{Glasser2019}. Quantum models of stochastic processes can have a lower memory dimension than any classical model of equal accuracy~\cite{Elliott2020,Yang2025}. This memory-dimension advantage has been demonstrated with a single photonic qubit~\cite{Wu2023} and, for models learned from discrete time series, against provably optimal classical models on quantum hardware~\cite{Yang2021}. Existing learning methods for HQMMs, however, maximize a likelihood over discrete symbols~\cite{Srinivasan2018,Adhikary2020}. No algorithm trains such a model exactly as a free-running generator of continuous, high-dimensional data, and no study has shown a memory-dimension advantage for generative models of such data, with controls that attribute it to the coherence of the memory.
	
	Here we introduce the coherent memory Born machine (CoMB) and the algorithms that make it trainable, compilable and testable. Our contributions are the following.
	\begin{enumerate}
		\item \emph{A generation algorithm with a fixed register.} CoMB produces a sample of $d$ values in $d$ rounds of a register of $\nm$ memory and $\nw$ work qubits, independent of $d$ (Algorithm~\ref{alg:sample}). Two design criteria, nestedness and depth scaling, make families with growing memory comparable (Sec.~\ref{sec:model}).
		\item \emph{An exact training algorithm.} A tilted-instrument identity expresses the characteristic function, the $k$-point laws and the path laws of the generated distribution as products of $d$ maps on the memory. It yields exact maximum mean discrepancy (MMD) losses and gradients at a cost linear in $d$, with no sampling of the model and no likelihood (Algorithm~\ref{alg:train}), and admits directional spectral components that resolve chosen observables (Sec.~\ref{sec:train}).
		\item \emph{Compilation and noise.} Every round compiles onto a single work qubit, and device noise folds exactly into the training loss (Secs.~\ref{sec:compile} and~\ref{sec:noise}).
		\item \emph{A controlled measure of coherence.} The memory sets the copula of the generated distribution and the readout its marginals, and the cut ranks of coherent and dephased memories are bounded. Every model is compared with its dephased twin and with the most general classical hidden Markov model of equal memory dimension, and the equivalent classical memory $\chi_{\rm equiv}$ measures what a coherent memory is worth (Sec.~\ref{sec:structure}).
		\item \emph{Benchmarks.} We evaluate the algorithm on two benchmarks chosen to differ in every respect: high-dimensional continuous data, complete $368$-cell calorimeter showers of CaloChallenge Dataset~1~\cite{calochallenge,ds1} generated from five qubits, and discrete sequential data, handwritten digits written as pen strokes~\cite{UCIdigits} by circuits of three qubits. We also execute the first rounds on an IBM Heron processor (Sec.~\ref{sec:bench}).
	\end{enumerate}
	
	These results advance generative quantum machine learning in three respects. They remove the scaling of the register with the data dimension, which has confined gate-based generators to compressed or patched representations. They replace sample-based loss estimates with exact losses and gradients, which removes shot noise from training and does not rely on probabilities a device cannot provide. And they provide a falsifiable protocol for attributing generative performance to quantum coherence, together with its first quantitative outcome on real data. The central empirical result is a memory-dimension advantage for learned generative models of real data, and it rests on three facts. First, the model's performance depends on its quantumness. Removing the coherence of its memory, with everything else fixed, raises the held-out loss of the four-level model by $23$ to $31\%$ at every training budget, in contrast to many quantum models in a recent systematic benchmark~\cite{Bowles2024}. Second, four coherent levels do the work of about six classical states, consistently on two unrelated benchmarks, and of at least five with $95\%$ confidence on showers at every training budget. Third, this holds when every model is trained to convergence, with the classical memories trained by the same loss and optimizer as the quantum ones. With eight levels, where all models approach the readout floor, coherent and classical memories become indistinguishable. Section~\ref{sec:discussion} discusses the scope of these claims. The contribution is a scalable model class, exact algorithms to train and compile it, and a measurement of what quantum coherence contributes to it.
	
	Section~\ref{sec:model} defines the model and its generation algorithm, Sec.~\ref{sec:train} derives the exact training algorithm, and Secs.~\ref{sec:structure}--\ref{sec:noise} establish its structural properties, the one-qubit compilation and the exact treatment of noise. Section~\ref{sec:bench} reports the benchmarks and the hardware execution, Sec.~\ref{sec:prior} discusses related work, and Sec.~\ref{sec:discussion} summarizes implications and limitations. Proofs, numerical checks and protocols are collected in the appendices.
	
	\section{The CoMB algorithm}\label{sec:model}
	
	\subsection{Register, rounds and instruments}
	
	\begin{definition}[CoMB]\label{def:comb}
		A coherent memory Born machine with $\nm$ memory qubits, $\nw$ work qubits and $d$ rounds is specified by unitaries $W_\beta$ on the memory and $V_\beta$ on memory and work, $\beta=1,\dots,d$. In round $\beta$ the register evolves under $U_\beta=V_\beta(W_\beta\otimes\Id)$, the work register is measured in the computational basis, yielding the token $c_\beta\in\{0,\dots,K-1\}$ with $K=2^{\nw}$, and is reset to $\ket{0}$. The memory, of dimension $\chi=2^{\nm}$, starts in the state $\rho_0=\ket{0}\bra{0}$ and is never measured.
	\end{definition}
	
	The memory interacts with the work register in every round and is conditioned on every measurement outcome. It therefore carries the dependence between rounds (Fig.~\ref{fig:chain}). Formally, round $\beta$ acts on the memory as the quantum instrument with Kraus operators
	\begin{equation}
		M_{\beta,c}=\big(\Id_\chi\otimes\bra{c}\big)\,U_\beta\,\big(\Id_\chi\otimes\ket{0}\big),\qquad \Lam_{\beta,c}(\rho)=M_{\beta,c}\,\rho\,M_{\beta,c}^\dagger .
		\label{eq:kraus}
	\end{equation}
	
	\begin{figure}[t]
		\centering
		\begin{quantikz}[column sep=6pt,row sep={0.62cm,between origins}]
			\lstick[label style={color=blue!60!black}]{$m_0\ \ket{0}$} & \qw\gategroup[2,steps=9,style={fill=blue!8,draw=none,inner ysep=1pt},background]{} & \gate[5,style={fill=cyan!15}]{U_1} & \qw & \qw & \gate[5,style={fill=cyan!15}]{U_2} & \qw & \qw & \gate[5,style={fill=cyan!15}]{U_3} & \qw & \rstick[2]{$\mathrm{Tr}_{\mathrm m}$} \\
			\lstick[label style={color=blue!60!black}]{$m_1\ \ket{0}$} & \qw & & \qw & \qw & & \qw & \qw & & \qw & \qw \\
			\lstick[label style={color=orange!75!black}]{$w_0\ \ket{0}$} & \qw\gategroup[3,steps=9,style={fill=orange!10,draw=none,inner ysep=1pt},background]{} & & \meter[style={fill=red!18}]{c_1} & \push{\textcolor{gray}{\ket{0}}\,} & & \meter[style={fill=red!18}]{c_2} & \push{\textcolor{gray}{\ket{0}}\,} & & \meter[style={fill=red!18}]{c_3} & \qw \\
			\lstick[label style={color=orange!75!black}]{$w_1\ \ket{0}$} & \qw & & \meter[style={fill=red!18}]{} & \push{\textcolor{gray}{\ket{0}}\,} & & \meter[style={fill=red!18}]{} & \push{\textcolor{gray}{\ket{0}}\,} & & \meter[style={fill=red!18}]{} & \qw \\
			\lstick[label style={color=orange!75!black}]{$w_2\ \ket{0}$} & \qw & & \meter[style={fill=red!18}]{} & \push{\textcolor{gray}{\ket{0}}\,} & & \meter[style={fill=red!18}]{} & \push{\textcolor{gray}{\ket{0}}\,} & & \meter[style={fill=red!18}]{} & \qw
		\end{quantikz}
		\caption{One sample at the level of wires, three rounds shown, qubit order $[m_0,m_1,w_0,w_1,w_2]$. The memory register (blue) is never measured and carries the dependence between rounds; the work register (orange) is measured after each round, yielding the token $c_\beta$, and reset.}
		\label{fig:chain}
	\end{figure}
	
	\begin{proposition}[Count-resolved instrument]\label{prop:instrument}
		The tokens of a sample have the joint distribution $P(c_1,\dots,c_d)=\Tr\bigl[\Lam_{d,c_d}\cdots\Lam_{1,c_1}(\rho_0)\bigr]$, and $\sum_c\Lam_{\beta,c}$ is completely positive and trace preserving for every $\beta$.
	\end{proposition}
	
	The memory state is propagated as a density matrix, not as a probability vector over basis states. This distinguishes a coherent memory from a classical one, and it is the property that the controls of Sec.~\ref{sec:controls} isolate.
	
	\subsection{Readout and decoder}
	
	Continuous data require a map from tokens to values. Coordinate $\beta$ is generated as $z_\beta=g_\beta(c_\beta+u_\beta)$, where $u_\beta$ is uniform on $[0,1)$ and $g_\beta$ is a strictly increasing piecewise linear readout that maps $[k,k+1)$ onto the $k$-th of $K$ bins with learned widths. Conditionally on the tokens, the coordinates are therefore independent and uniform within their bins. A fixed linear decoder $y=Az+b$ maps the coordinates to the data space. For the showers, $z$ are principal components and $y$ cell energies. For discrete data, such as the pen moves of Sec.~\ref{sec:digits}, the tokens are the data and no readout is required. On a device, one execution of the circuit returns one token sequence and hence one sample. No probabilities are estimated from repeated shots and no classical sampling step follows the measurement. Algorithm~\ref{alg:sample} summarizes the generation of one sample.
	
	\begin{figure}[t]
		\algtitle{Generating one sample with CoMB}\label{alg:sample}
		\begin{algorithmic}[1]
			\Require round unitaries $W_\beta$, $V_\beta$; readouts $g_\beta$; decoder $(A,b)$
			\State prepare the memory and work registers in $\ket{0}$
			\For{$\beta=1,\dots,d$}
			\State apply $W_\beta$ to the memory and $V_\beta$ to memory and work
			\State measure the work register, giving the token $c_\beta$, and reset it to $\ket{0}$
			\State draw $u_\beta$ uniformly from $[0,1)$ and set $z_\beta\gets g_\beta(c_\beta+u_\beta)$
			\EndFor
			\State \Return $y\gets Az+b$
		\end{algorithmic}
		\algend
	\end{figure}
	
	\subsection{Model classes and controls}\label{sec:controls}
	
	We consider four model classes with identical readout, decoder and training.
	
	\begin{definition}[Controls]\label{def:controls}
		(i) The \emph{circuit CoMB} realizes $W_\beta$ and $V_\beta$ as hardware-efficient circuits of $L_w$ and $L_v$ layers, each layer consisting of learned $R_Y$ and $R_Z$ rotations on every qubit followed by controlled-$Z$ gates along a line (Fig.~\ref{fig:block}). (ii) Its \emph{dephased twin} has identical parameters and dephases the memory completely in the computational basis after every round. (iii) The \emph{classical hidden Markov model} (HMM) with $\chi$ states has, in every round, a column-stochastic transition matrix and a stochastic emission matrix parameterized by softmax functions. (iv) The \emph{general coherent memory} replaces the circuit by an arbitrary isometry from the memory to memory, work and an optional environment of dimension $E$.
	\end{definition}
	
	The twin differs from the circuit CoMB only in the coherence of the memory and therefore measures the contribution of coherence at fixed parameters. The HMM is the most general classical memory of dimension $\chi$ and therefore the relevant benchmark at equal memory size. The general coherent memory is the limit of unlimited circuit depth. With an environment of dimension $E\ge\chi^2$ it contains the HMM (Proposition~\ref{prop:contain}). Without an environment, the memory's state conditioned on the tokens remains pure, and the model is a Born machine on a matrix product state of bond dimension $\chi$. At equal bond dimension, Born machines and hidden Markov models are incomparable, each representing distributions that the other needs a larger dimension for~\cite{Glasser2019}. This applies to every circuit in this paper and to the general coherent memory used for the showers, so their comparison with the classical HMM of equal dimension is an empirical question, not a consequence of containment.
	
	\begin{figure}[b]
		\centering
		\resizebox{\columnwidth}{!}{%
			\begin{quantikz}[column sep=7pt,row sep={0.72cm,between origins}]
				\lstick[label style={color=blue!60!black}]{$m_0$} & \gate[style={fill=cyan!15}]{R_Y(\theta)}\gategroup[2,steps=3,style={dashed,rounded corners,inner xsep=2pt},background,label style={anchor=south,yshift=0.05cm}]{$W_\beta$} & \gate[style={fill=cyan!15}]{R_Z(\theta)} & \ctrl{1} & \qw & \gate[style={fill=cyan!15}]{R_Y(\theta)}\gategroup[5,steps=5,style={dashed,rounded corners,inner xsep=2pt},background,label style={anchor=south,yshift=0.05cm}]{$V_\beta$} & \gate[style={fill=cyan!15}]{R_Z(\theta)} & \ctrl{2} & \qw & \qw & \qw & \rstick{$\cdots$}\qw \\
				\lstick[label style={color=blue!60!black}]{$m_1$} & \gate[style={fill=cyan!15}]{R_Y(\theta)} & \gate[style={fill=cyan!15}]{R_Z(\theta)} & \control{} & \qw & \gate[style={fill=cyan!15}]{R_Y(\theta)} & \gate[style={fill=cyan!15}]{R_Z(\theta)} & \qw & \ctrl{1} & \ctrl{2} & \qw & \rstick{$\cdots$}\qw \\
				\lstick[label style={color=orange!75!black}]{$w_0$} & \qw & \qw & \qw & \qw & \gate[style={fill=cyan!15}]{R_Y(\theta)} & \gate[style={fill=cyan!15}]{R_Z(\theta)} & \control{} & \control{} & \qw & \meter[style={fill=red!18}]{} & \rstick[wires=3]{$c_\beta$} \\
				\lstick[label style={color=orange!75!black}]{$w_1$} & \qw & \qw & \qw & \qw & \gate[style={fill=cyan!15}]{R_Y(\theta)} & \gate[style={fill=cyan!15}]{R_Z(\theta)} & \ctrl{1} & \qw & \control{} & \meter[style={fill=red!18}]{} & \\
				\lstick[label style={color=orange!75!black}]{$w_2$} & \qw & \qw & \qw & \qw & \gate[style={fill=cyan!15}]{R_Y(\theta)} & \gate[style={fill=cyan!15}]{R_Z(\theta)} & \control{} & \qw & \qw & \meter[style={fill=red!18}]{} &
		\end{quantikz}}
		\caption{The unitary of one round, $U_\beta=V_\beta(W_\beta\otimes\mathbb{I})$, for $\nm=2$ and $\nw=3$, with one of the $L_w=2$ layers of $W_\beta$ and one of the $L_v=5$ layers of $V_\beta$. Each layer applies learned $R_Y$ and $R_Z$ rotations to every qubit, followed by controlled-$Z$ gates along the line $m_0$--$w_0$--$m_1$--$w_1$--$w_2$.}
		\label{fig:block}
	\end{figure}
	
	\begin{proposition}[Containment]\label{prop:contain}
		If $E\ge\chi^2$, the general coherent memory with $\chi$ levels and $K$ tokens contains every classical $\chi$-state HMM with $K$ tokens.
	\end{proposition}
	
	\subsection{Nestedness and depth scaling}\label{sec:nest}
	
	Comparisons across memory sizes require that enlarging the memory extends the model family rather than altering it. We formalize this requirement and a companion criterion for the circuit depth.
	
	\begin{definition}[Nestedness]\label{def:nest}
		A family of circuit ansätze indexed by $\nm$ is \emph{nested} if, for every parameter setting of the ansatz with $\nm$ memory qubits, there is a parameter setting of the ansatz with $\nm+1$ memory qubits that generates the same distribution.
	\end{definition}
	
	For two memory qubits we use the line $m_0$--$w_0$--$m_1$--$w_1$--$w_2$, which alternates the registers and maps onto neighbouring qubits of a heavy-hexagon lattice. Extending the alternation to a third memory qubit places it between $w_1$ and $w_2$ and removes their coupling, which violates nestedness, because $w_2$ decouples when the new qubit is idle. Appending every memory qubit beyond the second after the work register, $m_0$--$w_0$--$m_1$--$w_1$--$w_2$--$m_2$, restores it.
	
	\begin{proposition}[Nested wiring]\label{prop:nest}
		The appended wiring is nested. The circuit with $\nm+1$ memory qubits, with zero rotation angles on the added qubit, generates exactly the distribution of the circuit with $\nm$ memory qubits. The alternating wiring is not nested for $\nm\ge3$.
	\end{proposition}
	
	Nestedness guarantees that the optimum cannot deteriorate with the memory, but it does not ensure that a circuit of fixed depth can exploit the added dimension. One round of the general coherent memory has $2\chi^2K-\chi^2$ real parameters, $240$ for $\chi=4$ and $960$ for $\chi=8$ with $K=8$. We therefore characterize an ansatz by its coverage $\rho_L$, the ratio of its angles per round to this number. Five layers give $\rho_5=58/240\approx0.24$ at $\chi=4$ but $\rho_5=72/960\approx0.08$ at $\chi=8$. Comparable coverage at $\chi=8$ requires $L_v=18$, $\rho_{18}=228/960\approx0.24$. Section~\ref{sec:matched} shows that both criteria are necessary for a coherent advantage at $\chi=8$.
	
	\section{Exact training}\label{sec:train}
	
	\subsection{Tilted-instrument identity}
	
	The central technical result is a factorization of tilted expectations over token sequences, the sequential analogue of the generating function of full counting statistics~\cite{levitov1996,Esposito2009,landi2024}.
	
	\begin{theorem}\label{thm:tilted}
		Let $w_t:\{0,\dots,K-1\}\to\mathbb{C}$ be arbitrary weights, one function per round. Then
		\begin{equation}
			\E\Big[\prod_{t=1}^d w_t(c_t)\Big]=\Tr\big[\mathcal{A}_d\circ\cdots\circ\mathcal{A}_1(\rho_0)\big],
			\label{eq:tilted}
		\end{equation}
		with the tilted maps $\mathcal{A}_t=\sum_{c}w_t(c)\,\Lambda_{t,c}$.
	\end{theorem}
	
	The right-hand side requires $d$ tilted maps, each a weighted sum of $K$ maps on the $\chi^2$-dimensional space of memory operators. Applied in Kraus form, they cost $O(dK\chi^3)$ operations, whereas the left-hand side is a sum over $K^d$ sequences. Every quantity used for training in this paper is an instance of Eq.~\eqref{eq:tilted} for a suitable choice of weights.
	
	\subsection{Characteristic function and MMD}
	
	Fourier weights yield the characteristic function. Let $w_{\beta,c}(t)=\E[e^{\,itz_\beta}\mid c_\beta=c]$ denote the characteristic function of the readout of round $\beta$ given token $c$. For a bin with centre $m_{\beta,c}$ and width $\delta_{\beta,c}$ it equals $e^{\,itm_{\beta,c}}\operatorname{sinc}(t\delta_{\beta,c}/2)$.
	
	\begin{proposition}[Exact characteristic function]\label{prop:cf}
		The characteristic function of the generated coordinates $z_\beta=g_\beta(c_\beta+u_\beta)$ is
		\begin{equation}
			\Phi(\omega)=\Tr\Big[\mathcal{A}_d(\omega_d)\cdots\mathcal{A}_1(\omega_1)\,(\rho_0)\Big],
			\label{eq:cf}
		\end{equation}
		with $\mathcal{A}_\beta(t)=\sum_c w_{\beta,c}(t)\,\Lam_{\beta,c}$. With the fixed decoder $y=Az+b$, the characteristic function of $y$ is $\Phi_y(\omega)=e^{i\omega\cdot b}\,\Phi(A^{T}\omega)$.
	\end{proposition}
	
	The characteristic function determines the distribution, and squared distances between characteristic functions are maximum mean discrepancies. By Bochner's theorem, a translation-invariant positive definite kernel is the Fourier transform of a nonnegative spectral measure $\mu$, and its two-sample statistic~\cite{Gretton2012} admits a spectral representation~\cite{sriperumbudur2010}.
	
	\begin{corollary}[Exact free-running MMD]\label{cor:mmd}
		For a translation-invariant kernel with $k(x,x)=1$, $k(x,x')=\int e^{i\omega\cdot(x-x')}\,\mu(d\omega)$, the identity $\MMD_k^2(P,Q_\theta)=\E_{\omega\sim\mu}|\Phi_P(\omega)-\Phi_\theta(\omega)|^2$ holds, the training loss below with frequencies drawn from $\mu$ is unbiased for $\MMD_k^2(\hat P_N,Q_\theta)$, and its gradient is unbiased for $\nabla_\theta\MMD_k^2(P,Q_\theta)$.
	\end{corollary}
	
	We use the MMD rather than the likelihood because only the MMD scales to hardware. The likelihood requires the model's probability of each training sample. A device returns samples, not probabilities, and the probability of a particular sequence of $d$ tokens, which decreases exponentially with $d$, would have to be estimated from a number of shots of the order of its inverse. The MMD requires only samples on a device, with a number of shots that does not grow with $d$, and in simulation it is given exactly by Proposition~\ref{prop:cf}. Our simulations could evaluate the likelihood only because memories of dimension four and eight are classically simulable.
	
	The training loss averages $|\hat\Phi_N(\omega_r)-\Phi_\theta(\omega_r)|^2$ over $R=256$ frequencies per step, drawn from a fixed pool of orthogonal random features~\cite{rahimi2007,yu2016} of a mixture of Gaussian kernels with bandwidths $\sigma/2$, $\sigma$ and $2\sigma$, where $\sigma$ is the median inter-sample distance. The model term is exact, so the only stochasticity is the choice of frequencies. The model is never sampled, and the loss trains the free-running generator rather than a one-step predictor conditioned on the true past. Because the pool is drawn once, the loss is unbiased for the MMD of the pool's random-feature kernel, which approximates the Gaussian mixture with an error of order $n_{\rm pool}^{-1/2}$ for a pool of $n_{\rm pool}$ frequencies.
	
	\paragraph{Directional spectral components.} Isotropic kernels with bandwidth of the order of the inter-sample distance are insensitive to the shape of one-dimensional projections in high dimension. Along any direction $a$, their spectral measure concentrates at frequencies far below $1/\sigma_a$, where $\sigma_a$ is the spread of the data along $a$, so that the loss constrains only the mean and the width of the projected distribution. Proposition~\ref{prop:cf} removes this limitation at no additional cost, because the characteristic function of $a\cdot y$ is $\Phi_y$ evaluated along $a$. Augmenting the spectral measure with frequencies $t\,a$, $|t|=r/\sigma_a$, adds a kernel that resolves the distribution of $a\cdot y$ to a fraction $1/r$ of its width, and the loss remains an exact MMD because a sum of kernels is a kernel. For the showers we include directions along the energy fractions of the two dense calorimeter layers (Sec.~\ref{sec:ds1}).
	
	\subsection{$k$-point and path MMD}
	
	Indicator weights yield the laws of token windows and token pairs. For a positive definite kernel $G$ on tokens, the sequence kernel
	\begin{multline}
		\mathcal{K}(c,c')=\sum_{k=1}^{3}\frac{\lambda_k}{d-k+1}\sum_t\prod_{j<k}G[c_{t+j},c'_{t+j}]\\
		+\frac{\lambda_p}{|\mathcal{P}|}\sum_{(t,s)\in\mathcal{P}}G[c_t,c'_t]\,G[c_s,c'_s]
	\end{multline}
	induces an MMD that is a sum of quadratic forms in the model--data differences of the window laws $W_k(t)$ and the pair laws $P_{ts}$, and vanishes if and only if all of them agree. Each law is an instance of Eq.~\eqref{eq:tilted}. For a pair,
	\begin{equation}
		P_{ts}(a,b)=\Tr\big[\Lambda_{s,b}\,T_{s-1}\cdots T_{t+1}\,\Lambda_{t,a}(r_{t-1})\big],
	\end{equation}
	with $T_u=\sum_c\Lambda_{u,c}$ and $r_{t-1}$ the memory state before round $t$. Each term is normalized by its value for a model without dependence between tokens. Quantities that accumulate over a sample are constrained by the \emph{path MMD}. If token $c$ carries an increment $v(c)$ and $X_\tau=\sum_{t<\tau}v(c_t)$, the joint characteristic function of $X_{\tau_1},\dots,X_{\tau_m}$ is Eq.~\eqref{eq:tilted} with $w_t(c)=\exp\big(i\,v(c)\cdot\sum_{j:\tau_j>t}\omega_j\big)$, although it involves correlations of all orders between tokens.
	
	\subsection{Gradients and complexity}
	
	Gradients follow from Eq.~\eqref{eq:tilted} by reverse-mode differentiation. A forward pass stores the memory state after each round. A backward pass propagates an adjoint vector through the transposed tilted maps, and the outer product of adjoint and stored state at round $\beta$ is the gradient with respect to $\mathcal{A}_\beta$, from which the chain rule proceeds through the Kraus operators into the circuit parameters. One training step therefore costs $O(R\,d\,K\chi^3)$ operations and $O(R\,d\,\chi^2)$ memory, independently of $K^d$. All quantities were verified against independent references (Appendix~\ref{app:checks}). Algorithm~\ref{alg:train} summarizes one training step.
	
	\begin{figure}[t]
		\algtitle{One exact training step}\label{alg:train}
		\begin{algorithmic}[1]
			\Require Kraus operators $M_{\beta,c}$ and readout bins (depending on $\theta$); decoder $(A,b)$; frequencies $\omega_1,\dots,\omega_R$ from the pool; data values $\hat\Phi_N(\omega_r)$
			\For{$r=1,\dots,R$}
			\State $\nu\gets A^{T}\omega_r$; \ $\sigma_0\gets\rho_0$
			\For{$\beta=1,\dots,d$}
			\State $\sigma_\beta\gets\sum_c w_{\beta,c}(\nu_\beta)\,M_{\beta,c}\,\sigma_{\beta-1}\,M_{\beta,c}^\dagger$ \Comment{tilted map, Eq.~\eqref{eq:cf}}
			\EndFor
			\State $\Phi_\theta(\omega_r)\gets e^{i\omega_r\cdot b}\,\Tr\,\sigma_d$
			\EndFor
			\State $L\gets\frac{1}{R}\sum_r\big|\hat\Phi_N(\omega_r)-\Phi_\theta(\omega_r)\big|^2$
			\State propagate the adjoint of $L$ from $\sigma_d$ back to $\sigma_0$ through the transposed tilted maps, accumulating $\partial L/\partial M_{\beta,c}$ and $\partial L/\partial w_{\beta,c}$ from the stored $\sigma_{\beta-1}$
			\State apply the chain rule to the circuit angles and the readout parameters
			\State \Return $L$ and $\nabla_\theta L$ \Comment{cost $O(R\,d\,K\chi^3)$}
		\end{algorithmic}
		\algend
	\end{figure}
	
	\section{Structure of the generated distribution}\label{sec:structure}
	
	The generated distribution separates into a dependence structure, set by the memory, and marginal distributions, set by the readout.
	
	\begin{proposition}[Copula--marginal factorization]\label{prop:copula}
		Let $t_\beta=c_\beta+u_\beta$. For any readouts $g_\beta$ that are continuous and strictly increasing, the copula of $z=(g_\beta(t_\beta))_\beta$ equals the checkerboard copula of the count vector $c$ and does not depend on $g$. The marginal distribution function of $z_\beta$ is $F_\beta(\zeta)=\sum_{c<k}p_\beta(c)+p_\beta(k)(\tau-k)$ with $\tau=g_\beta^{-1}(\zeta)\in[k,k+1)$. For any continuous target marginal $F^\star_\beta$ and any count marginal $p_\beta$ with full support, the quantile readout $\ell_{\beta,c}=F_\beta^{\star-1}\bigl(\sum_{c'<c}p_\beta(c')\bigr)$, $0<c<K$, with end points enclosing the support, reproduces $F^\star_\beta$ on the grid of the bin edges.
	\end{proposition}
	
	Proposition~\ref{prop:copula} has two practical consequences~\cite{nelsen2006,genest2007}. The readout can be initialized, or reset, at the data quantiles without altering the dependence structure learned by the memory. Conversely, the density of a coordinate equals the probability of a bin divided by its width, so that a readout bin that contracts while retaining its probability produces a spurious narrow peak. The directional spectral components of Sec.~\ref{sec:train} are designed to penalize exactly such defects in the observables that matter.
	
	\begin{proposition}[Memory structure]\label{prop:rank}
		(i) If the memory starts in a diagonal state and is completely dephased after every round, the model is a $\chi$-state hidden Markov model with transition--emission probabilities $|\bra{i'}M_{\beta,x}\ket{i}|^2$. (ii) For every cut $1,\dots,k\,|\,k+1,\dots,d$, the matrix of joint count probabilities $P_{(c_{\le k}),(c_{>k})}$ factorizes as $P_{p,f}=\Tr(E_fB_p)$ with positive semidefinite $\chi\times\chi$ matrices $B_p$ and $E_f$, so its PSD rank is at most $\chi$. For the dephased model, its nonnegative rank is at most $\chi$.
	\end{proposition}
	
	Because the positive semidefinite rank of a nonnegative matrix can be much smaller than its nonnegative rank~\cite{fawzi2015}, Proposition~\ref{prop:rank} identifies the representational resource of a coherent memory. It does not determine its value on given data, which depends on the extent to which the data's dependence structure requires PSD factorizations. Rank estimates from empirical cut matrices are unsuitable for this purpose, since with realistic sample sizes they are dominated by sampling noise~\cite{comb1}. We therefore define an operational measure.
	
	\begin{definition}[Equivalent classical memory]\label{def:equiv}
		For a coherent model with memory dimension $\chi$, the equivalent classical memory $\chi_{\rm equiv}$ is the number of states at which classical HMMs reach the held-out loss of the coherent model. The classical HMMs are trained with the same loss and protocol, and their loss is interpolated log-linearly in the number of states.
	\end{definition}
	
	A value $\chi_{\rm equiv}=\chi$ indicates that coherence confers no representational benefit on the data at hand. A value $\chi_{\rm equiv}=\chi^2$ would correspond to the maximal separation compatible with Proposition~\ref{prop:rank}.
	
	\section{Compilation onto one work qubit}\label{sec:compile}
	
	The number of work qubits can be reduced to one without changing the generated distribution, by implementing each $K$-outcome instrument as a binary tree of two-outcome instruments~\cite{Andersson2008,Ivashkov2024}.
	
	\begin{theorem}\label{thm:tree}
		Let $\{M_o\}$, $o\in\{0,1\}^n$, be the Kraus operators of one round, with $\sum_oM_o^\dagger M_o=\Id$ and $K=2^n$. For every prefix $p$ let $E_p=\sum_{o\supseteq p}M_o^\dagger M_o$. The node operators
		\begin{equation}
			N_{p,b}=V_{pb}\,E_{pb}^{1/2}E_p^{-1/2},\qquad b\in\{0,1\},
		\end{equation}
		with $V=\Id$ at inner nodes and $V_o$ the polar unitary of $M_o$ at the leaves, satisfy $N_{p_{n-1},b_n}\cdots N_{(),b_1}=M_o$ for every $o$. Every node is a two-outcome instrument, $\sum_bN_{p,b}^\dagger N_{p,b}=\Id$ on the support of $E_p$.
	\end{theorem}
	
	Because the memory undergoes the same instrument, Theorem~\ref{thm:tilted} applies unchanged and the compiled model is trained with the same losses. A CoMB with two memory qubits thus runs on three qubits with mid-circuit measurement, reset and classical feed-forward (Fig.~\ref{fig:tree}). In practice the tree is parameterized directly by one circuit per node and trained as such.
	
	\begin{figure*}[t]
		\centering
		\begin{minipage}[t]{0.285\textwidth}
			\centering{\small (a) one round of CoMB}\par\medskip
			\begin{quantikz}[column sep=5pt,row sep={0.6cm,between origins}]
				\lstick[label style={color=blue!60!black}]{$m_0$} & \gate[2,style={fill=blue!8}]{W_t} & \gate[5,style={fill=gray!12}]{V_t} & \qw & \qw \\
				\lstick[label style={color=blue!60!black}]{$m_1$} & & & \qw & \qw \\
				\lstick[label style={color=orange!75!black}]{$w_0\ \ket{0}$} & \qw & & \meter[style={fill=red!18}]{} & \cw \rstick{$b_1$} \\
				\lstick[label style={color=orange!75!black}]{$w_1\ \ket{0}$} & \qw & & \meter[style={fill=red!18}]{} & \cw \rstick{$b_2$} \\
				\lstick[label style={color=orange!75!black}]{$w_2\ \ket{0}$} & \qw & & \meter[style={fill=red!18}]{} & \cw \rstick{$b_3$}
			\end{quantikz}
		\end{minipage}%
		\begin{minipage}[t]{0.07\textwidth}
			\vspace*{1.45cm}\centering\tikz{\draw[-{Stealth[length=2.2mm]},very thick,gray] (0,0) -- (2,0) node[midway,above,font=\scriptsize,black] {Theorem~\ref{thm:tree}};}
		\end{minipage}%
		\begin{minipage}[t]{0.645\textwidth}
			\centering{\small (b) the same round, compiled onto one work qubit}\par\medskip
			\begin{quantikz}[column sep=4pt,row sep={0.8cm,between origins}]
				\lstick[label style={color=blue!60!black}]{$m_0$} & \gate[2,style={fill=blue!8}]{W_t} & \gate[3,style={fill=gray!12}]{N_{()}} & \qw & \qw & \gate[3,style={fill=gray!12}]{N_{(b_1)}} & \qw & \qw & \gate[3,style={fill=gray!12}]{N_{(b_1b_2)}} & \qw & \qw \\
				\lstick[label style={color=blue!60!black}]{$m_1$} & & & \qw & \qw & & \qw & \qw & & \qw & \qw \\
				\lstick[label style={color=orange!75!black}]{$w\ \ket{0}$} & \qw & & \meter[style={fill=red!18}]{b_1} & \push{\textcolor{gray}{\ket{0}}\,} & & \meter[style={fill=red!18}]{b_2} & \push{\textcolor{gray}{\ket{0}}\,} & & \meter[style={fill=red!18}]{b_3}& \qw
			\end{quantikz}
		\end{minipage}
		\caption{Compilation of a round onto one work qubit (Theorem~\ref{thm:tree}) for the three-qubit work register of Figs.~\ref{fig:chain} and~\ref{fig:block}. (a) The round with the three work qubits measured jointly, yielding $c_t=b_1b_2b_3$. (b) The equivalent instrument on a single work qubit: node circuits emit the bits sequentially, the qubit is measured and reset after every node, and classical feed-forward selects the next node. Both generate the same distribution.}
		\label{fig:tree}
	\end{figure*}
	
	\section{Exact evaluation under device noise}\label{sec:noise}
	
	Theorem~\ref{thm:tilted} holds for arbitrary completely positive instruments. A device noise model can therefore be composed into the instruments of each round, and the loss of the deployed circuit evaluated exactly, without shots and without hardware access. We include depolarizing noise after every gate, imperfect reset of the work register, readout errors, and amplitude damping and dephasing of the memory during each measure-and-reset window. In a measurement tree, a readout error alters both the recorded bit and the node selected by feed-forward. Indexing the noisy instruments by the recorded outcome accounts for both effects exactly. We use a trapped-ion and a superconducting device model with published error rates (Appendix~\ref{app:protocols}). Scaling the superconducting model by a factor $\lambda$ defines the threshold $\lambda^\star$ at which a coherent model ceases to outperform its dephased twin, a hardware requirement that can be checked directly against calibration data.
	
	\section{Benchmarks}\label{sec:bench}
	
	We evaluate CoMB on two benchmarks chosen to differ in every respect that matters for a sequential generator. The first is high-dimensional and continuous: each sample has $369$ real coordinates with strong, long-range dependence. The second is discrete and sequential: each sample is a sequence of $56$ tokens from an alphabet of $16$, whose structure accumulates over the whole sequence. On both, every CoMB is trained next to its dephased twin and the most general classical HMM of equal memory dimension, with identical readout, loss, data and training budget, over several seeds, and all results are measured on held-out data (Sec.~\ref{sec:controls}). Between the two benchmarks, we report an execution of the shower model on quantum hardware.
	
	\subsection{High-dimensional continuous data: calorimeter showers}\label{sec:ds1}
	
	\paragraph{Data and protocol.}
	
	We use the $65$~GeV photon showers of CaloChallenge Dataset~1~\cite{calochallenge,ds1}, simulated with \textsc{Geant4} in a calorimeter with five sampled layers of $8$, $160$, $190$, $5$ and $5$ cells. Each shower is represented by its total energy and its $368$ energy fractions, standardized and rotated into principal components. The model generates the $369$ components, and the fixed decoder recovers the cells. The data are split into $4\,000$ training showers, $2\,000$ for model selection, $2\,000$ held out for the reported loss, and $1\,200$ test showers for the physics metrics. All CoMB models use $\nw=3$. The loss is the characteristic-function MMD with $20\%$ of its spectral weight assigned to directional components along the energy fractions of the two dense layers, at $r=2$ and $4$. Without these components, training with the isotropic kernel produces a double-peaked energy distribution in both dense layers that the isotropic loss does not penalize, because the leading principal component, which governs the energy split between these layers, develops two contracted readout bins (Proposition~\ref{prop:copula}). The classical HMM is trained with the identical loss. Readouts are initialized by the quantile rule and circuit angles from $\mathcal{N}(0,0.6^2)$. We train with Adam~\cite{Kingma2015} for $3\,000$ steps, with peak learning rates of $0.1$ for the angles and $0.03$ for the readouts, linear warm-up over $50$ steps and cosine decay, and select the step with the lowest exact validation MMD$^2$, evaluated every $25$ steps. Every configuration is trained with $30$ independent seeds.
	
	\paragraph{Generation at full resolution.}
	\begin{table*}[tb]
		\centering
		\caption{Calorimeter showers of CaloChallenge Dataset~1 ($368$ cells). All models are trained for $3\,000$ steps with the loss of Sec.~\ref{sec:train} including directional components along the two dense layers; values are means over $30$ seeds. Rows marked $^{\ast}$ use three memory qubits with nested wiring and $18$ layers (Sec.~\ref{sec:matched}). MMD$^2$: held-out loss on $2\,000$ showers excluded from training and selection, with the standard error over seeds. Remaining columns are computed against $1\,200$ \textsc{Geant4} test showers: per-cell Wasserstein-1 distance in units of the cell's standard deviation for dense and sparse cells (W1); fraction of sparse cells at zero (\textsc{Geant4}: $21\%$); mean absolute difference of cell--cell rank correlations (Spearman); captured share of the correlation structure between the memoryless model ($0\%$) and the finite-sample floor ($100\%$) ($\kappa$); median absolute deviation of the total energy relative to \textsc{Geant4} ($E$ spread). The QFAN row is from Ref.~\cite{qfan}.}
		\label{tab:main}
		\small\setlength{\tabcolsep}{3pt}
		\begin{tabular}{lccccccccc}
			\toprule
			Model & Qubits & Params & MMD$^2$ ($10^{-4}$, test) & W1 dense & W1 sparse & Zeros & Spearman & $\kappa$ & $E$ spread \\
			\midrule
			CoMB $\chi{=}4$ & 5 & $24\,354$ & $2.07\pm0.06$ & $0.119$ & $0.219$ & $22.9\%$ & $0.051$ & $84.6\%$ & $1.09$ \\
			Dephased twin $\chi{=}4$ & -- & $24\,354$ & $2.70\pm0.06$ & $0.120$ & $0.219$ & $23.1\%$ & $0.050$ & $85.0\%$ & $1.09$ \\
			HMM $\chi{=}4$ & -- & $20\,664$ & $2.81\pm0.09$ & $0.122$ & $0.218$ & $23.3\%$ & $0.048$ & $87.2\%$ & $1.04$ \\
			\midrule
			CoMB $\chi{=}8^{\ast}$ & 6 & $87\,084$ & $1.77\pm0.03$ & $0.119$ & $0.216$ & $22.7\%$ & $0.051$ & $84.5\%$ & $1.08$ \\
			Dephased twin $\chi{=}8^{\ast}$ & -- & $87\,084$ & $2.45\pm0.05$ & $0.123$ & $0.218$ & $22.8\%$ & $0.052$ & $83.9\%$ & $1.07$ \\
			HMM $\chi{=}8$ & -- & $50\,184$ & $1.86\pm0.02$ & $0.121$ & $0.217$ & $23.0\%$ & $0.049$ & $86.4\%$ & $1.07$ \\
			QFAN~\cite{qfan} & 4 & -- & -- & $0.276$ & $0.381$ & $3.8\%$ & $0.105$ & $48\%$ & $0.62$ \\
			No correlations & -- & -- & -- & $0.023$ & $0.012$ & $20.7\%$ & $0.159$ & $0\%$ & $1.00$ \\
			Floor (train vs test) & -- & -- & -- & $0.048$ & $0.048$ & $21.0\%$ & $0.031$ & $100\%$ & $0.96$ \\
			\bottomrule
		\end{tabular}
	\end{table*}
	
\begin{figure*}[tb]
		\centering
		\subfloat[]{\figfile{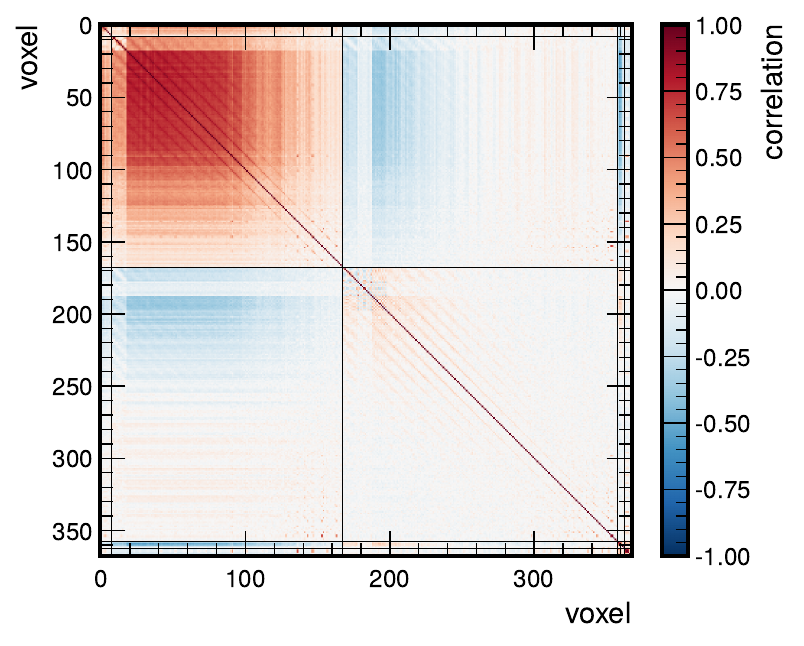}{.6\columnwidth}}\hspace{0.6em}
		\subfloat[]{\figfile{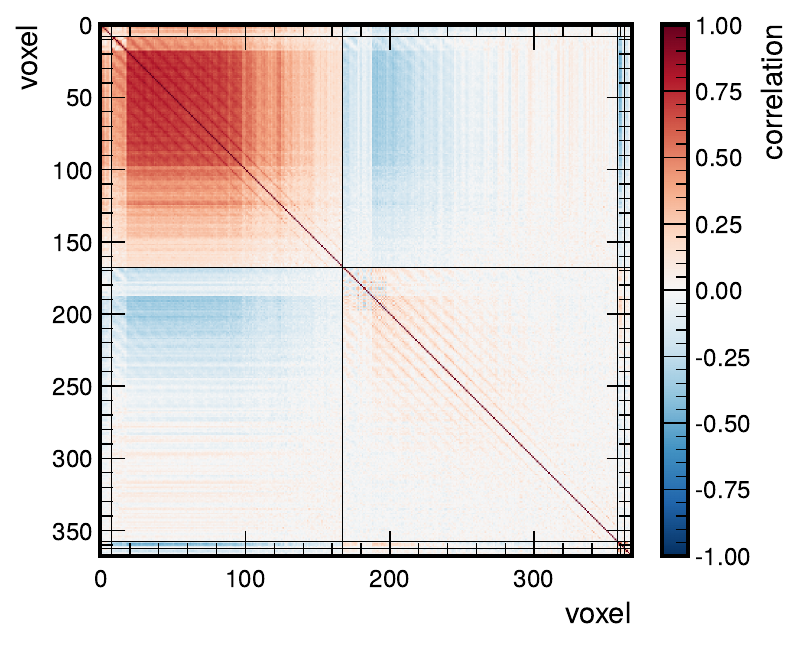}{.6\columnwidth}}
		\subfloat[]{\figfile{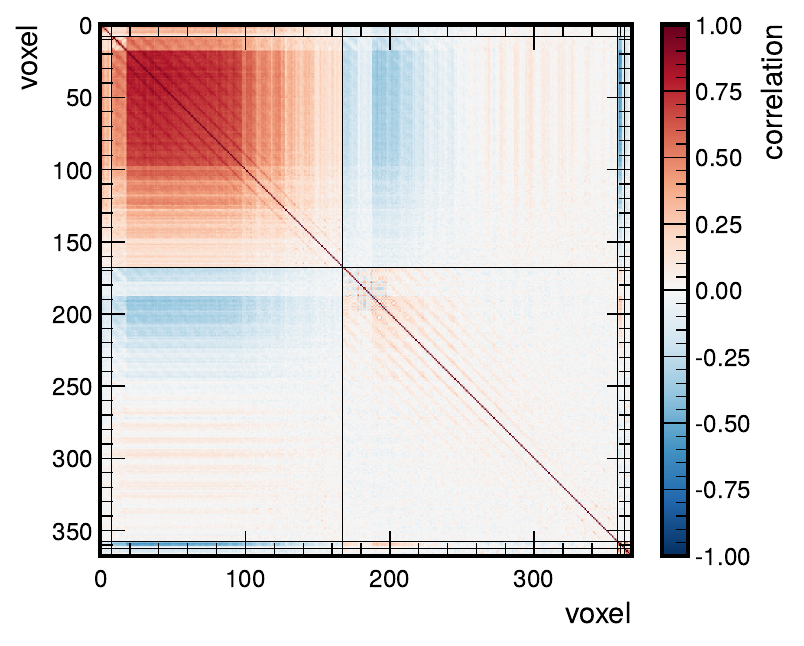}{.6\columnwidth}}
		\caption{Cell--cell correlation matrices of the $368$-cell showers, ordered by layer and ring: (a) \textsc{Geant4} test showers, (b) CoMB and (c) the classical HMM with $\chi=4$ (representative runs, seeds $0$ and $1$). Lines separate the calorimeter layers.}
		\label{fig:corr}
	\end{figure*}

\begin{figure*}[tb]
		\centering
		\subfloat[]{\figfile{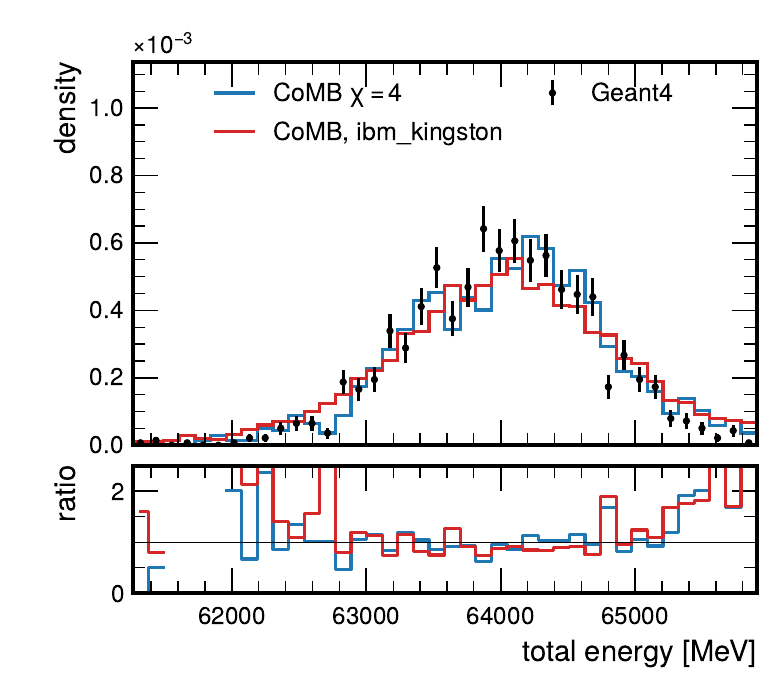}{0.33\textwidth}}
		\subfloat[]{\figfile{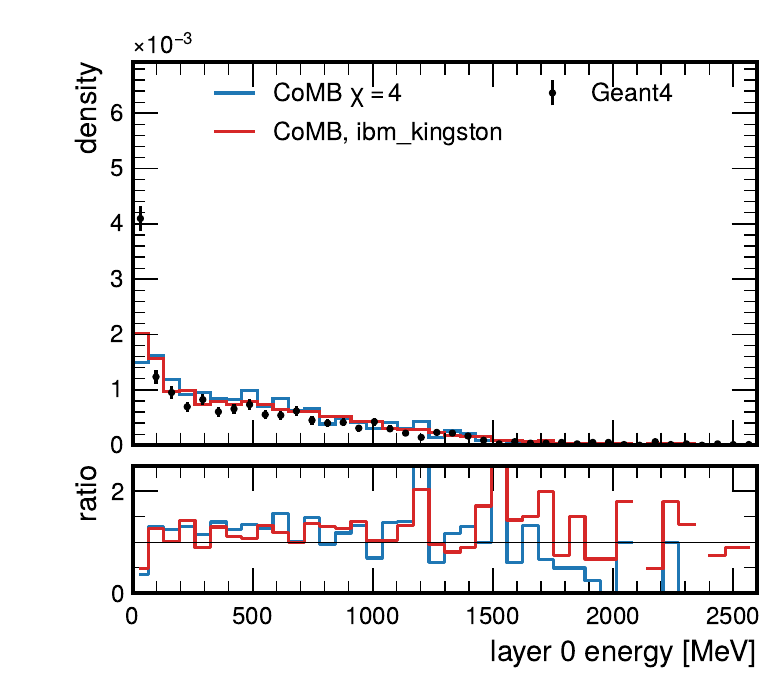}{0.33\textwidth}}
		\subfloat[]{\figfile{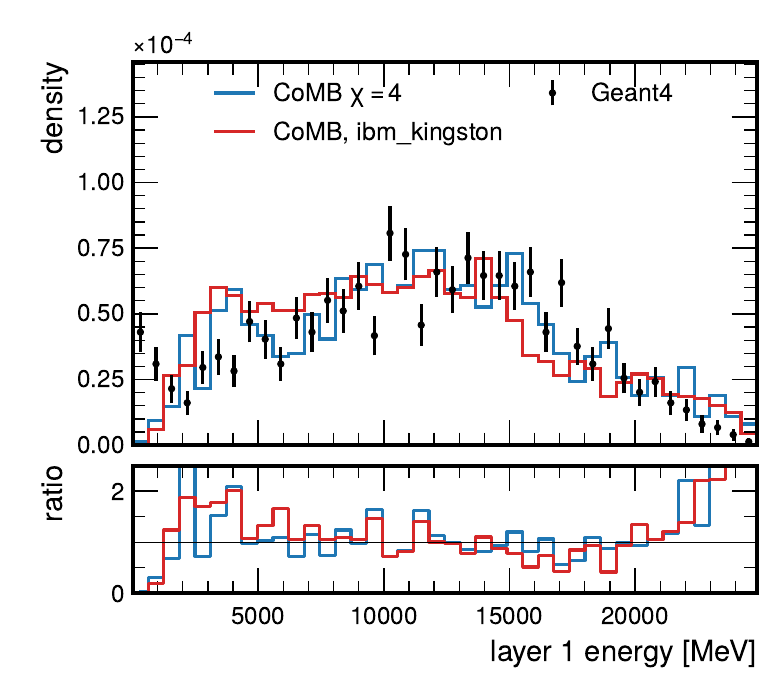}{0.33\textwidth}}\\
		\subfloat[]{\figfile{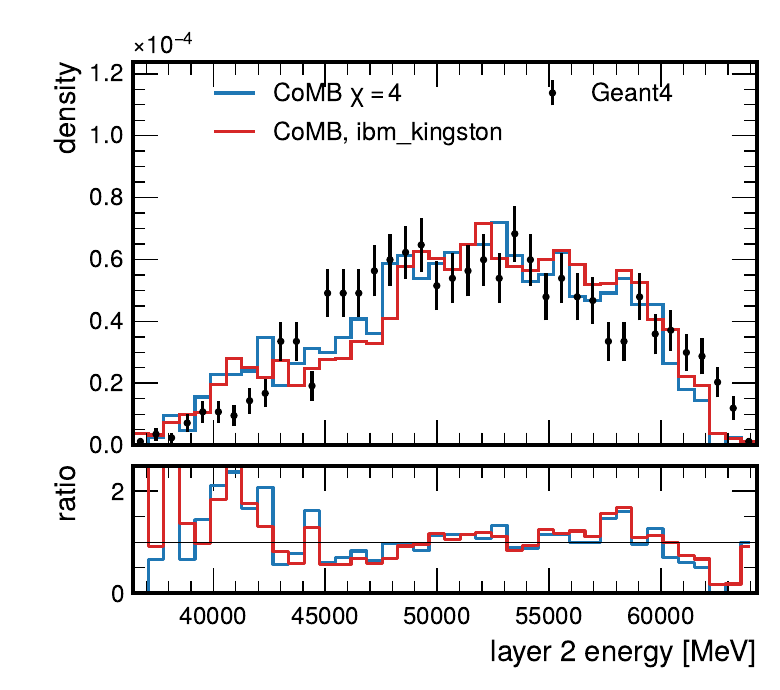}{0.33\textwidth}}
		\subfloat[]{\figfile{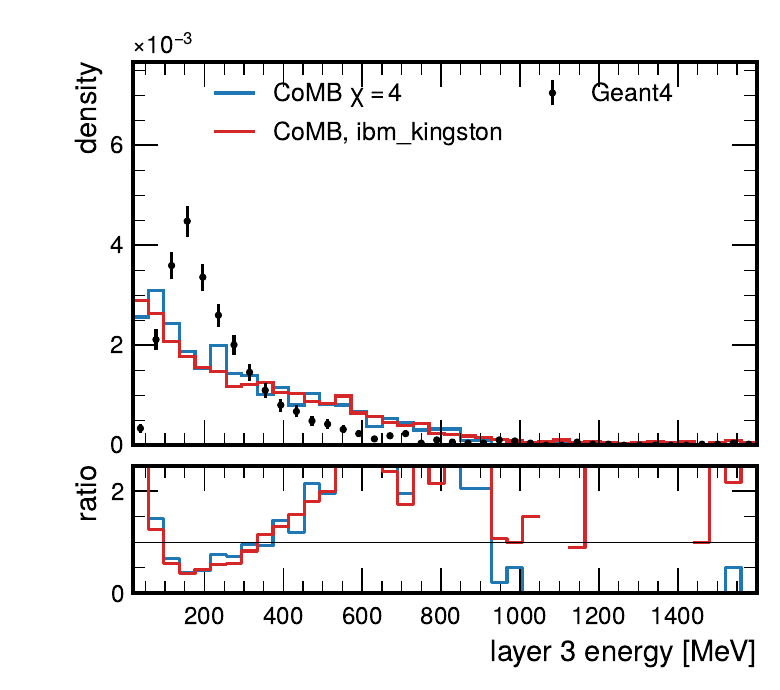}{0.33\textwidth}}
		\subfloat[]{\figfile{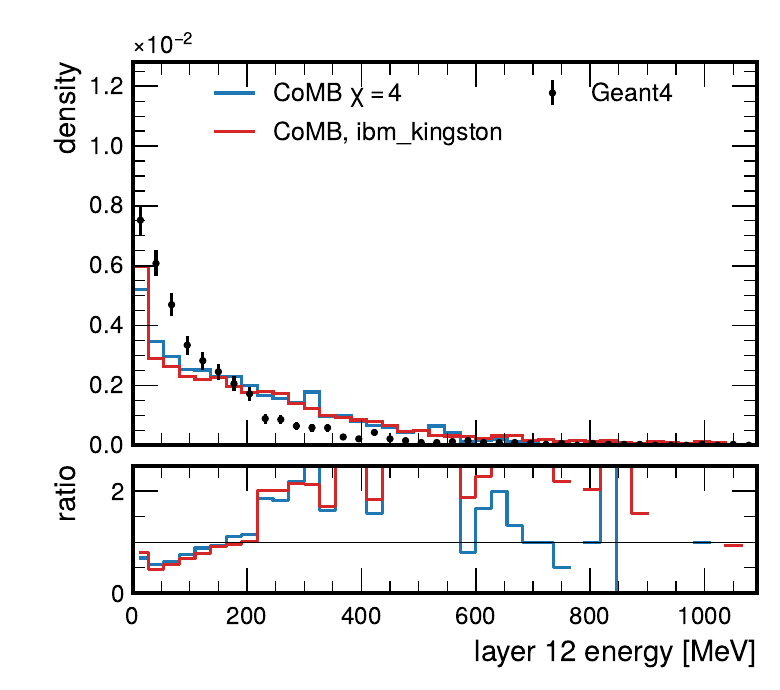}{0.33\textwidth}}
		\caption{(a) Total energy and energy deposited in (b) layer~0, (c) layer~1, (d) layer~2, (e) layer~3 and (f) layer~12, for \textsc{Geant4} (points), CoMB and the classical HMM with $\chi=4$; lower panels show the ratio to \textsc{Geant4}.}
		\label{fig:energies}
	\end{figure*}

\begin{figure*}[tb]
		\centering
		\subfloat[]{\figfile{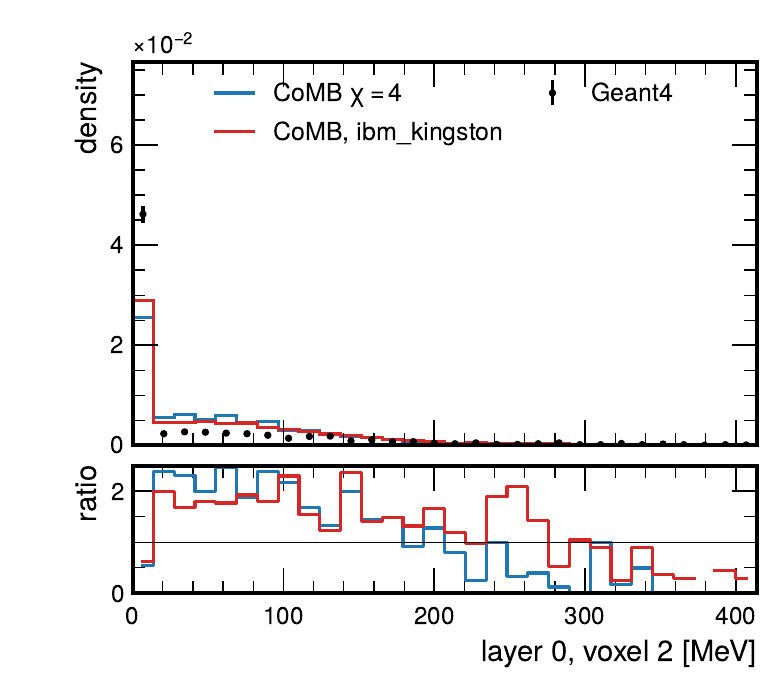}{0.33\textwidth}}
		\subfloat[]{\figfile{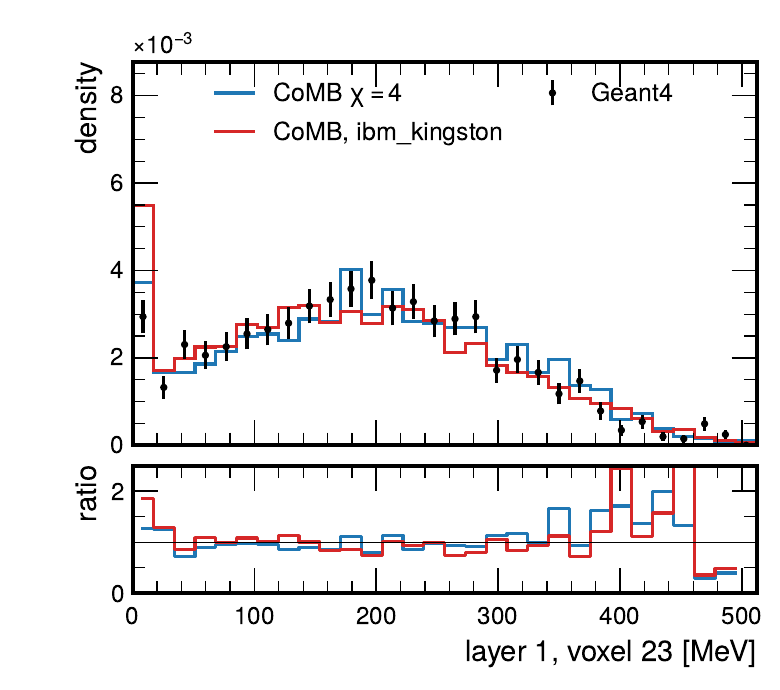}{0.33\textwidth}}
		\subfloat[]{\figfile{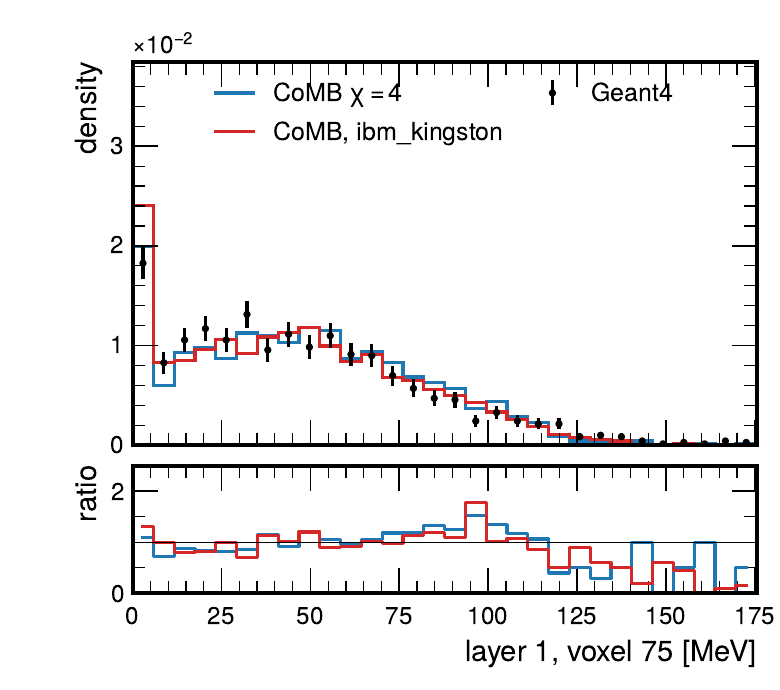}{0.33\textwidth}}\\
		\subfloat[]{\figfile{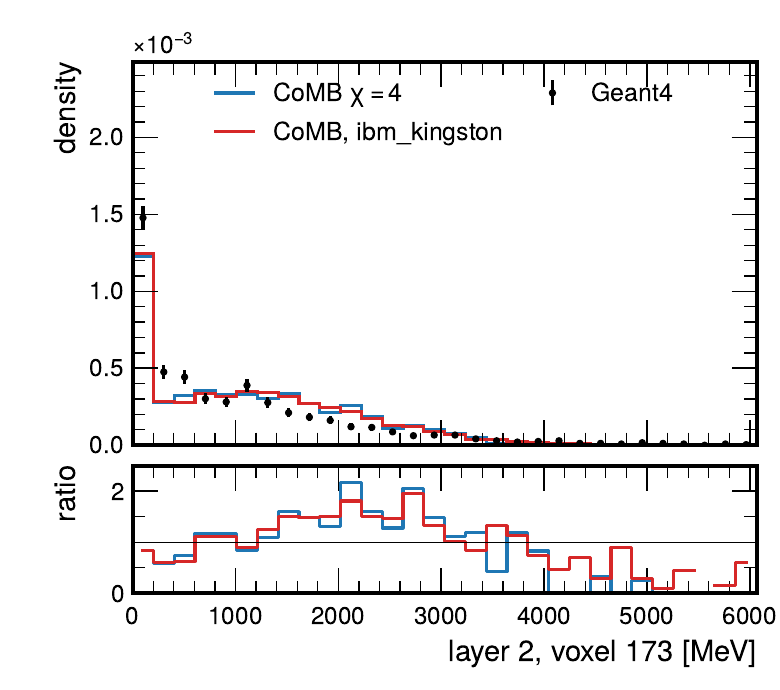}{0.33\textwidth}}
		\subfloat[]{\figfile{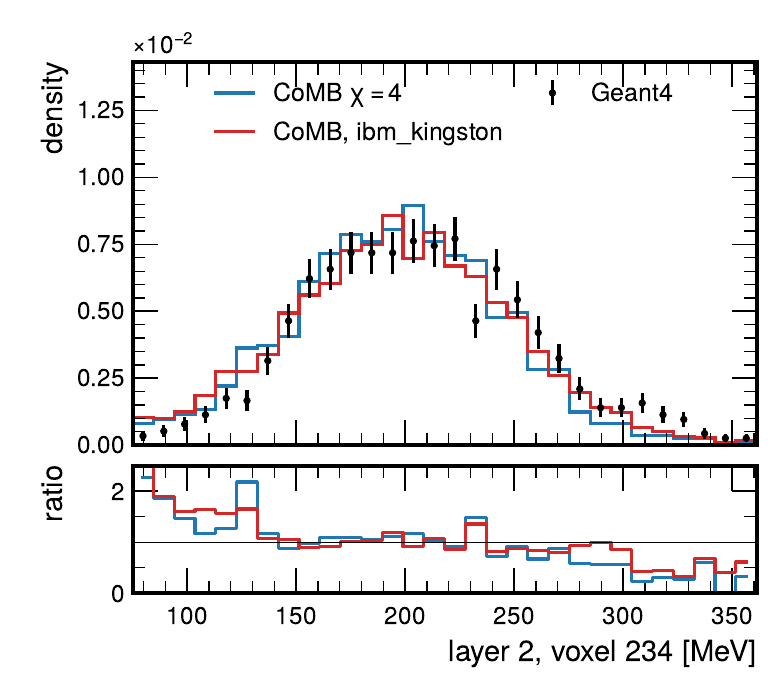}{0.33\textwidth}}
		\subfloat[]{\figfile{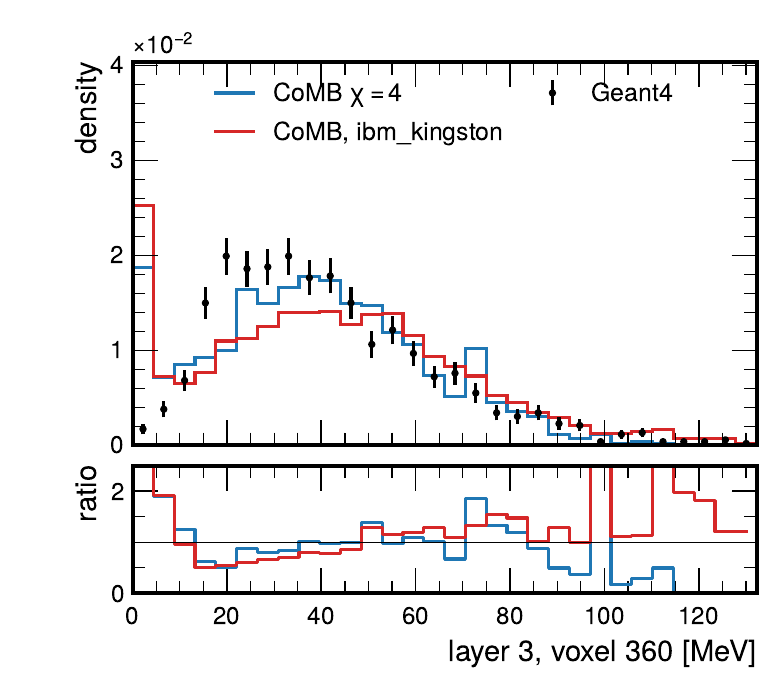}{0.32\textwidth}}
		\caption{Energy spectra of six cells, indexed ring by ring: (a) cell~2 of layer~0, (b) cells~23 and (c)~75 of layer~1, (d) cells~173 and (e)~234 of layer~2, (f) cell~360 of layer~3; \textsc{Geant4} (points), CoMB and the classical HMM, with ratio panels.}
		\label{fig:marg}
	\end{figure*}

\begin{figure*}[tb]
		\centering
		\figfile{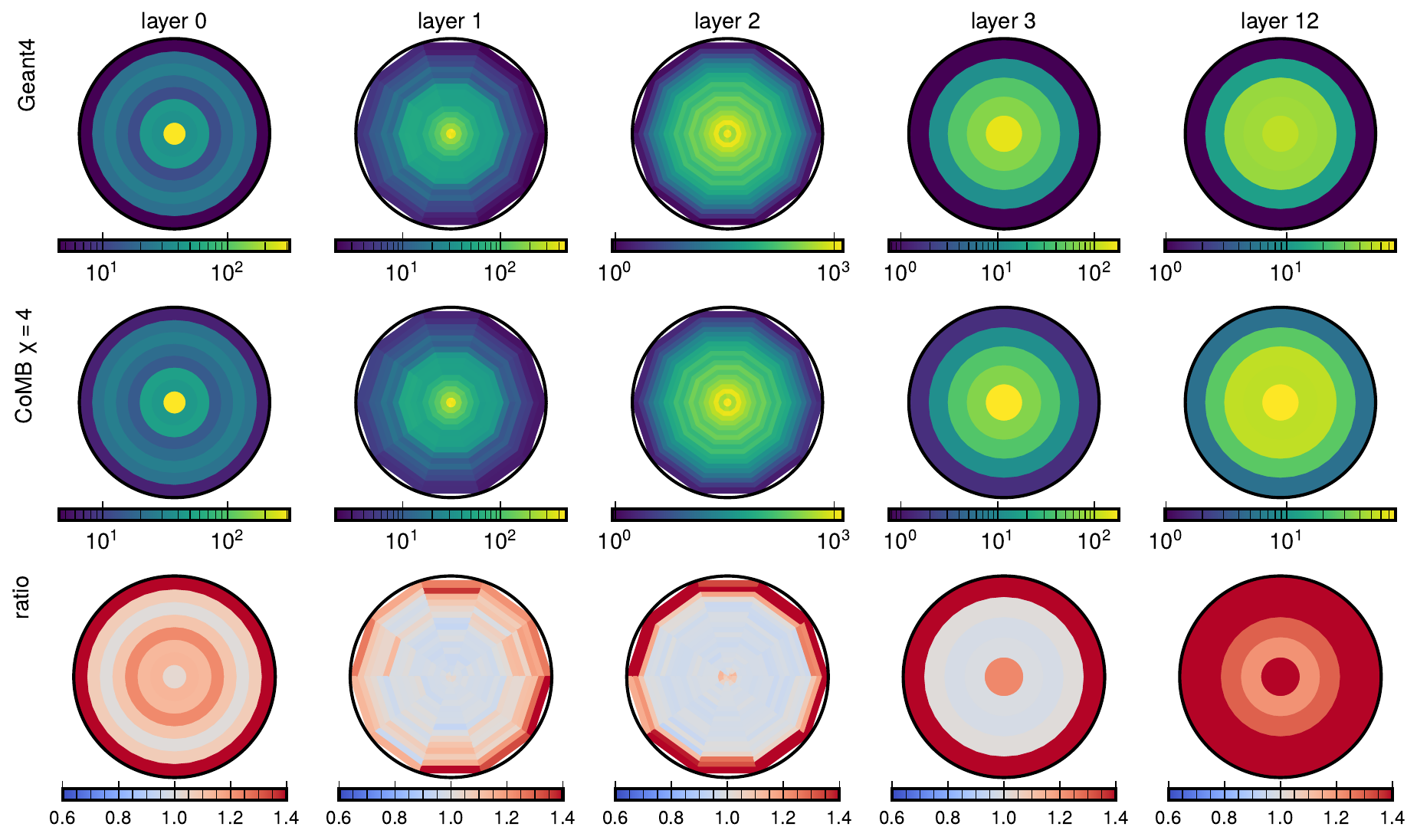}{\textwidth}
		\caption{Mean shower per layer as polar maps on a logarithmic scale, one scale per layer: \textsc{Geant4} (top), CoMB (middle) and their ratio (bottom). Rings are drawn with equal width.}
		\label{fig:polar}
	\end{figure*}

	Table~\ref{tab:main} and Figs.~\ref{fig:corr}, \ref{fig:energies}, \ref{fig:marg} and~\ref{fig:polar} characterize the generated showers. The energies of both dense layers are unimodal and follow \textsc{Geant4} (Fig.~\ref{fig:energies}). The cell--cell correlation matrix reproduces the correlated block of the first dense layer, the anticorrelation between the two dense layers across rings, and the ring structure of the second dense layer (Fig.~\ref{fig:corr}). The mean shower agrees with \textsc{Geant4} at the level of a few percent in the shower core, with deviations confined to the sparse outer rings and the thin layers (Fig.~\ref{fig:polar}). Quantitatively, the five-qubit CoMB captures $\kappa=84.6\%$ of the achievable correlation structure, attains per-cell Wasserstein distances of $0.119$ and $0.219$ standard deviations for dense and sparse cells, reproduces the fraction of empty sparse cells ($22.9\%$ against $21\%$) and, to within $10\%$, the shower-to-shower spread of the total energy. A gradient-boosted classifier still separates generated from simulated showers for every model in Table~\ref{tab:main}. The models are therefore not competitive with classical fast-simulation generators, and the comparisons that follow concern memories, not simulation fidelity.
	
	The classical HMM with four states, trained with the same readout and loss, reaches $\kappa=87.2\%$, compared with $48\%$ for the four-qubit generator of Ref.~\cite{qfan}, and reduces its dense-cell Wasserstein distance from $0.276$ to $0.122$. Most of the improvement over Ref.~\cite{qfan} is thus attributable to the readout and the exact training, which all models in Table~\ref{tab:main} share. The contribution of the memory is quantified separately below.
	
	\paragraph{Coherent versus classical memory.}\label{sec:matched} Every model was trained at three budgets, described in Appendix~\ref{app:protocols}: a fixed budget of $3\,000$ steps, a budget of $6\,000$ steps obtained by continuing each trained model, and training to convergence. Table~\ref{tab:budget} reports the four-level comparison at all three. At every budget the CoMB circuit outperforms the classical HMM of the same size, by $26\%$, $22\%$ and $15\%$, which corresponds to $7.2$, $6.4$ and $4.7$ standard errors. The margin shrinks as training becomes longer, but the equivalent classical memory barely changes, $\chi_{\rm equiv}=6.1$, $5.8$ and $5.8$ states, in agreement with the $6.2$ states found on digits (Sec.~\ref{sec:digits}). Resampling the seeds gives $95\%$ intervals of $5.5$ to $7.1$, $5.3$ to $6.8$ and $5.0$ to $7.2$ states at the three budgets, and every resample exceeds four states (Appendix~\ref{app:protocols}). The coherence gap does not shrink. The loss of CoMB lies $23\%$, $30\%$ and $31\%$ below that of its dephased twin. The general coherent memory without an environment ($E=1$), started from the trained circuits, does better still. At convergence it reaches $(1.17\pm0.01)\times10^{-4}$, which is $21\%$ below the classical HMM ($10.0$ standard errors), with $\chi_{\rm equiv}=7.7$. Because all models are evaluated on the same held-out showers, test-set fluctuations largely cancel in these differences, and the quoted errors reflect training variability. The advantage resides in the joint distribution. The summary metrics of Table~\ref{tab:main}, which probe single cells or pairs of cells, do not separate the memories, consistent with the observation that losses restricted to low-order statistics cannot distinguish quantum from classical generators~\cite{BornUltimatum2025}.
	
	\begin{table}[tb]
		\centering\footnotesize\setlength{\tabcolsep}{3pt}
		\begin{tabular}{lccc}
			\toprule
			$\chi=4$ & $3\,000$ steps & $6\,000$ steps & converged\\
			\midrule
			CoMB circuit & $2.07\pm0.06$ & $1.54\pm0.04$ & $1.26\pm0.04$\\
			dephased twin & $2.70\pm0.06$ & $2.20\pm0.06$ & $1.83\pm0.05$\\
			classical HMM & $2.81\pm0.09$ & $1.99\pm0.05$ & $1.48\pm0.03$\\
			CoMB vs.\ HMM & $+7.2$ & $+6.4$ & $+4.7$\\
			$\chi_{\rm equiv}$ & $6.1$ & $5.8$ & $5.8$\\
			\midrule
			general memory & -- & $1.39\pm0.02$ & $1.17\pm0.01$\\
			its dephased twin & -- & $2.00\pm0.05$ & $1.64\pm0.04$\\
			general vs.\ HMM & -- & $+10.2$ & $+10.0$\\
			$\chi_{\rm equiv}$ & -- & $7.7$ & $7.7$\\
			\midrule
			$\chi=8$ & & &\\
			CoMB circuit & $1.77\pm0.03$ & $1.37\pm0.01$ & $1.19\pm0.01$\\
			dephased twin & $2.45\pm0.05$ & $1.75\pm0.03$ & --\\
			classical HMM & $1.86\pm0.02$ & $1.37\pm0.02$ & $1.16\pm0.02$\\
			CoMB vs.\ HMM & $+2.6$ & $+0.1$ & $-1.4$\\
			$\chi_{\rm equiv}$ & $9.1$ & $8.0$ & $7.3$\\
			\bottomrule
		\end{tabular}
		\caption{Coherent and classical memories at three training budgets, $30$ seeds per model. The circuit with eight levels has nested wiring and $18$ layers, and its dephased twin was continued to $6\,000$ steps only. Held-out MMD$^2$ ($10^{-4}$) with standard errors, the difference between the coherent model and the classical HMM of the same size in standard errors (positive where the coherent model is better), and the equivalent classical memory, read from classical HMMs trained with the same budget. The general coherent memory is started from the trained circuits and therefore has no entry at $3\,000$ steps (Appendix~\ref{app:protocols}).}
		\label{tab:budget}
	\end{table}
	
\begin{figure}[tb]
		\centering
		\figfile{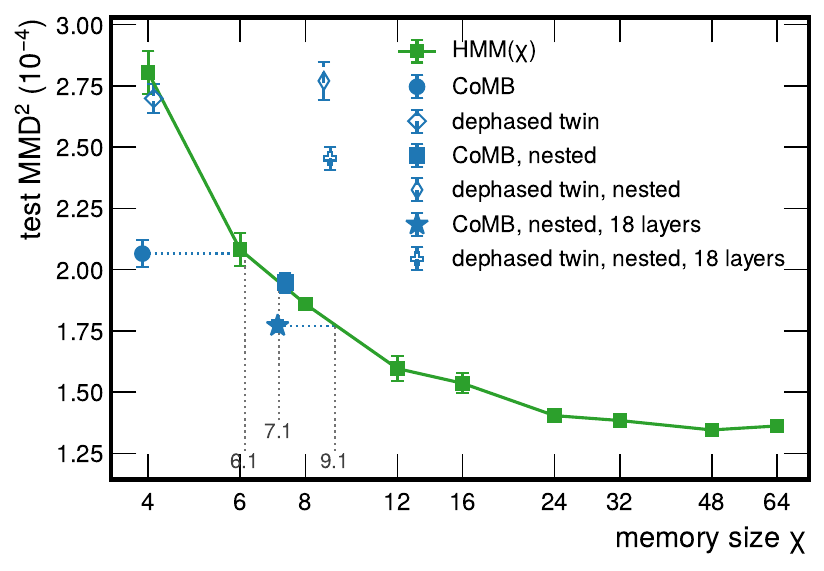}{\linewidth}
		\caption{Held-out MMD$^2$ against memory size after $3\,000$ training steps: classical HMMs with $4$ to $64$ states ($30$ seeds at $\chi=4$ and $8$, $10$ otherwise), and CoMB (filled) with its dephased twin (open) at $\chi=4$ and, with nested wiring, at $\chi=8$ with five layers (squares) and $18$ layers (stars), $30$ seeds each. Points are means, bars standard errors. Dotted lines indicate the equivalent classical memory of each CoMB configuration, $6.1$, $7.1$ and $9.1$.}
		\label{fig:ds1memory}
	\end{figure}

	\paragraph{Memory scaling.} Figure~\ref{fig:ds1memory} places every memory on the classical curve at the fixed budget of $3\,000$ steps. Table~\ref{tab:chi8} reports all three configurations evaluated at $\chi=8$. The alternating wiring, which violates nestedness (Proposition~\ref{prop:nest}), yields no improvement over $\chi=4$ and falls $9\%$ short of the classical HMM with eight states. The nested wiring at five layers, whose coverage is $\rho_5\approx0.08$, improves the loss, and its trained models occupy $7.2$ to $7.9$ of their eight levels and reach all $64$ independent memory states along sampled histories, yet it remains $5\%$ above the classical HMM. With nested wiring and $18$ layers, which restores the coverage of the four-level circuit, CoMB reaches $(1.77\pm0.03)\times10^{-4}$, which is $5\%$ below the classical HMM of equal size ($2.6$ standard errors) and $28\%$ below its dephased twin ($12.9$ standard errors), with $\chi_{\rm equiv}=9.1$. Both design criteria of Sec.~\ref{sec:nest} are therefore necessary for a coherent advantage of the circuit at $\chi=8$ at this budget. With longer training the circuit and the classical memory become indistinguishable (Table~\ref{tab:budget}). After $6\,000$ steps the circuit reaches $(1.370\pm0.010)\times10^{-4}$, against $(1.372\pm0.015)\times10^{-4}$ for the classical HMM ($0.1$ standard errors), and at convergence $(1.189\pm0.011)\times10^{-4}$, against $(1.162\pm0.015)\times10^{-4}$ ($-1.4$ standard errors), with $\chi_{\rm equiv}=8.0$ and $7.3$. At convergence its $95\%$ interval, $5.9$ to $8.3$ states, contains eight. It remains $22\%$ below its dephased twin after $6\,000$ steps ($12.1$ standard errors). The margin at the fixed budget therefore reflects how fast the models train, not what they can represent. Trained to convergence, the general coherent memory with eight levels reaches $(1.147\pm0.009)\times10^{-4}$, against $(1.162\pm0.015)\times10^{-4}$ for the classical HMM with eight states. That is a difference of $1\%$ ($0.9$ standard errors), with $\chi_{\rm equiv}=8.6$, while the coherent memory remains $28\%$ below its own dephased twin ($22.5$ standard errors). This environment-free memory is not guaranteed to contain the classical HMM (Sec.~\ref{sec:controls}), and its optimization is slower than that of the circuits. Its dephased version, which contains the dephased circuit, ends above it after $6\,000$ steps, at $1.87$ against $1.75\times10^{-4}$. The tie therefore bounds what this model reached, not what a coherent memory of this size can represent. Converged classical HMMs saturate from about $12$ states on, at $1.08$ to $1.10\times10^{-4}$, a floor imposed by the readout and decoder shared by all models. At eight levels all models lie close to this floor, which leaves little room for a memory advantage.
	
	\paragraph{Training budgets.} A fixed budget does not guarantee that every model reaches the best loss it can attain, and continuing the classical HMMs after $3\,000$ steps lowers their held-out loss by $26$ to $29\%$. Every comparison in this section is therefore reported with its budget. The four-level comparison is repeated after $6\,000$ steps and at convergence (Table~\ref{tab:budget}). The configurations with eight memory levels are compared at the fixed budget in Table~\ref{tab:chi8}, and the best of them is followed to $6\,000$ steps and to convergence (Table~\ref{tab:budget}).
	
	\begin{table}[tb]
		\centering\footnotesize\setlength{\tabcolsep}{2.5pt}
		\begin{tabular}{lccccc}
			\toprule
			$\chi=8$ & CoMB & twin & vs.\ twin & vs.\ HMM & $\chi_{\rm equiv}$\\
			\midrule
			altern., $L_v{=}5$ & $2.03\pm0.07$ & $2.93\pm0.09$ & $+8.1$ & $-2.3$ & $6.4$\\
			nested, $L_v{=}5$ & $1.95\pm0.04$ & $2.77\pm0.08$ & $+9.4$ & $-1.9$ & $7.1$\\
			nested, $L_v{=}18$ & $1.77\pm0.03$ & $2.45\pm0.05$ & $+12.9$ & $+2.6$ & $9.1$\\
			\bottomrule
		\end{tabular}
		\caption{All CoMB configurations evaluated with three memory qubits, $30$ seeds each, after $3\,000$ training steps: held-out MMD$^2$ ($10^{-4}$) of CoMB and its dephased twin, the differences to the twin and to the classical HMM with eight states ($(1.86\pm0.02)\times10^{-4}$) in standard errors, positive where CoMB is better, and the equivalent classical memory. A fourth configuration, alternating wiring with $18$ layers, was discontinued once the alternating wiring was found to violate nestedness, and has no result.}
		\label{tab:chi8}
	\end{table}
	
	\subsection{Execution on quantum hardware}\label{sec:hw}
	
	A complete shower requires a dynamic circuit of $369$ rounds, beyond present coherence times. The property that the architecture requires of hardware, however, can be tested on the first rounds. The memory must transport dependence between rounds through the measurements and resets of the work register. We executed the first twelve rounds of a trained five-qubit model, which generate the twelve leading principal components, as one dynamic circuit on \texttt{ibm\_kingston}, a $156$-qubit Heron r2 processor, with $4\,000$ shots (job \texttt{db34tb7r11fs7397r770}). The circuit comprises $264$ two-qubit gates before and $333$ after routing onto five neighbouring qubits. It was verified against the model instruments to $10^{-15}$ before submission, and a noisy simulation reproduced the exact count distributions within the multinomial floor. The executed model was trained with the isotropic kernel, before the directional components were introduced. It shares the architecture and register of the models in Table~\ref{tab:main}, so the experiment characterizes the architecture on this device.
	
\begin{figure}[tb]
		\centering
		\subfloat[]{\figfile{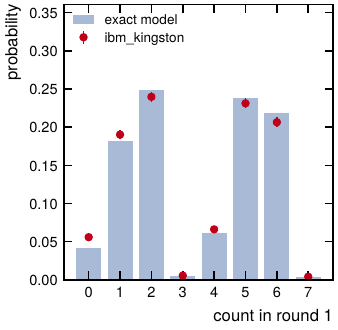}{0.41\textwidth}}\hspace{0.6em}
		\subfloat[]{\figfile{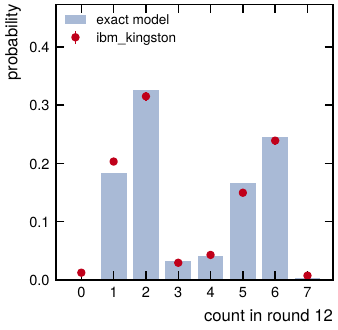}{0.41\textwidth}}
		\caption{Twelve rounds of a trained five-qubit model executed as one dynamic circuit on \texttt{ibm\_kingston} ($4\,000$ shots): token distributions of (a) the first and (b) the twelfth round, exact (bars) and measured (points with statistical errors), with their total-variation distances.}
		\label{fig:hw_tv}
	\end{figure}

	The total-variation distance between measured and exact token distributions is $0.028$ in the first round and $0.038$ in the twelfth, with a mean of $0.082$ against a shot-noise floor of $0.018$ (Fig.~\ref{fig:hw_tv}), and does not accumulate along the circuit. The mean absolute Spearman correlation between the tokens of different rounds is $0.047$ for the exact model and $0.031$ on the device, so that two thirds of the inter-round dependence is preserved. For the most strongly dependent pair, rounds nine and ten, the correlation is $-0.57$ in the model and $-0.27$ on the device. Since the work register is reset between rounds, this dependence is transported exclusively by the memory qubits. Each device record can be completed by the exact model from the memory state selected by its outcomes, which yields full showers whose leading coordinates are generated on the processor.
	
	\subsection{Discrete sequential data: handwritten digits}\label{sec:digits}
	
	Handwritten digits drawn as pen-stroke sequences provide a discrete test case that is structurally unrelated to calorimeter showers and admits an independent assessment of sample quality. Each image of the UCI optical digits~\cite{UCIdigits} ($\UN$ images of $8\times8$ pixels, upsampled to $28\times28$) is binarized, thinned to a one-pixel skeleton~\cite{ZhangSuen1984}, and traced by a deterministic walk that visits every skeleton pixel, preferring the neighbour with the fewest unvisited onward neighbours and the smallest turn. Pen-up moves connect disjoint strokes and pad sequences to length $\Ud$. Each move is one of $16$ tokens, eight directions with pen up or down, i.e.\ the outcomes of four work qubits. A classifier trained on rendered round trips identifies $\UCeil$ of held-out digits, the ceiling attainable in this representation. The training objective is the $k$-point MMD with windows of up to three moves and pairs at lags $3$ to $96$, plus the path MMD of the pen position every four moves with weight two (Sec.~\ref{sec:train}).
	
\begin{figure}[b]
		\centering
		\figfile{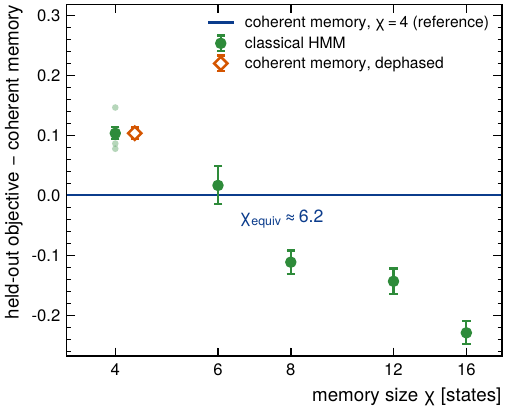}{\columnwidth}
		\caption{Held-out objective on the digits relative to the general coherent memory with $\chi=4$ (zero line; absolute value $\UCohMean$), against memory size: classical HMMs with $4$ to $16$ states and, at $\chi=4$, the dephased memory at identical parameters (open diamond). All models are evaluated on the same test digits; differences carry paired errors.}
		\label{fig:memory}
	\end{figure}

\begin{figure*}[t]
		\centering
		\subfloat[]{\figfile{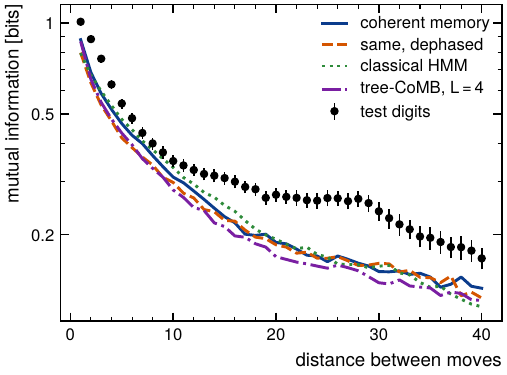}{.8\columnwidth}}
		\subfloat[]{\figfile{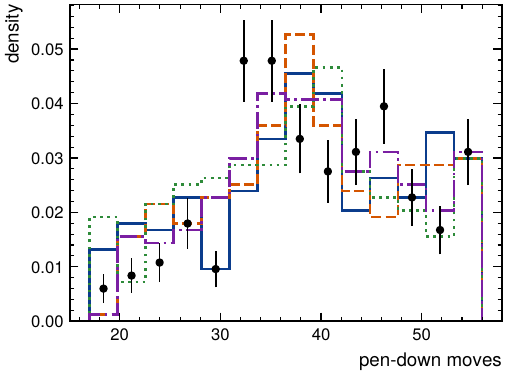}{.8\columnwidth}}\\
		\subfloat[]{\figfile{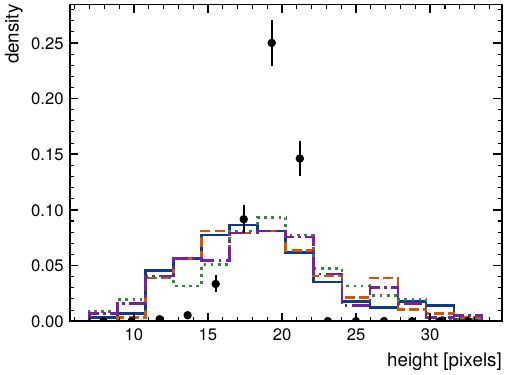}{.8\columnwidth}}
		\subfloat[]{\figfile{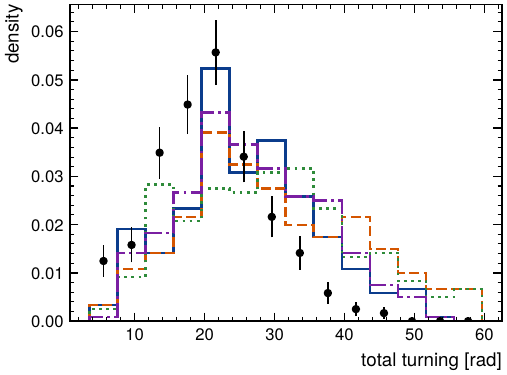}{.8\columnwidth}}
		\caption{Statistics of test digits (points, with statistical errors) and of the models with $\chi=4$ (lines). (a) Mutual information between two pen moves as a function of their separation, averaged over positions and estimated from equal numbers of sequences. (b)--(d) Distributions of the number of pen-down moves, of the height, and of the total turning of the digits.}
		\label{fig:statistics}
	\end{figure*}

	\paragraph{Coherence at fixed memory size.} At $\chi=4$ we compare the general coherent memory with $16$ environment levels, its dephased counterpart at identical parameters, and classical HMMs with $4$ to $16$ states (Fig.~\ref{fig:memory}). Over $\USeeds$ seeds, \UBeatSentence. The worst coherent run attains $\UCohWorst$ on the full held-out objective, against $\UOtherBest$ for the best of all other runs. Dephasing raises the objective by $\UGainDeph$, and the classical HMM of equal size is $\UGainHmm$ worse. As paired differences on the same test digits, with jackknife errors, the dephased memory lies $\UDiffDeph$ and the classical HMM $\UDiffHmm$ above the coherent memory. Classical HMMs reach the coherent four-level memory at $\chi_{\rm equiv}=\UChiEquiv$, in agreement with the value obtained on showers. A coherent memory without environment levels performs equivalently, so circuits of unlimited depth would require no environment qubits for this task. Figure~\ref{fig:statistics} relates the objective to directly interpretable statistics of the digits.
	
\begin{figure}[t]
		\centering
		\figfile{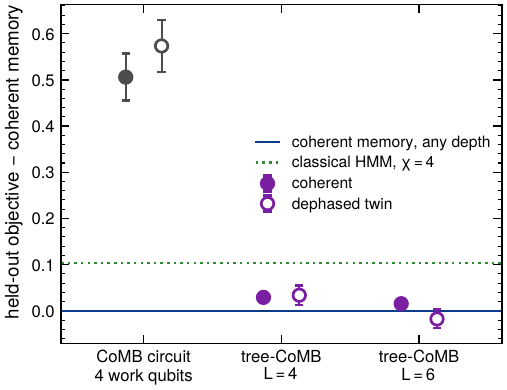}{\columnwidth}
		\caption{Circuits of fixed depth (filled) and their dephased twins (open), relative to the general coherent memory (zero line): the circuit CoMB with four work qubits and four layers, and the measurement-tree CoMB with one work qubit at node depths~4 and~6. Dotted: classical HMM with $\chi=4$. Error bars: jackknife errors of the differences.}
		\label{fig:circuits}
	\end{figure}

	\paragraph{Circuits of fixed depth.} Figure~\ref{fig:circuits} quantifies how much of the advantage of the general memory survives in circuits of fixed depth. On the full objective, the circuit with four work qubits outperforms its dephased twin ($\UDiffComb$, circuit minus twin), whereas the measurement-tree circuits do not differ significantly from theirs ($\UDiffTreeFour$ at node depth~4, $\UDiffTreeSix$ at depth~6), although they do on the local part of the objective (Table~\ref{tab:results}). As on showers, a circuit of fixed depth retains only part of the coherent advantage of the general memory.
	
\begin{figure*}[t]
		\centering
		\subfloat[]{\figfile{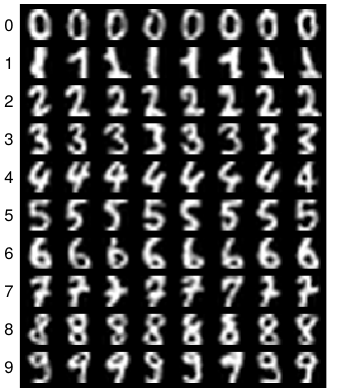}{0.31\textwidth}}\hspace{0.6em}\subfloat[]{\figfile{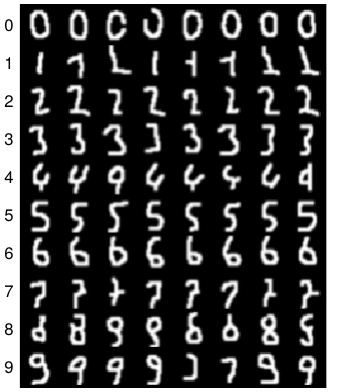}{0.31\textwidth}}\hspace{0.6em}\subfloat[]{\figfile{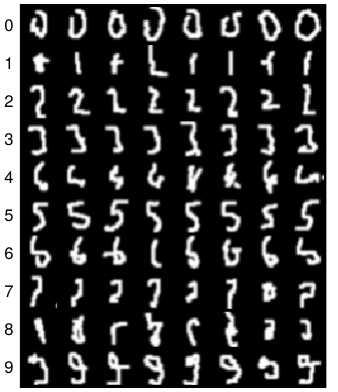}{0.31\textwidth}}
		\caption{(a) Held-out test digits, one row per class. (b) The same digits rendered from their pen-stroke sequences, the attainable ceiling in this representation. (c) Samples of the class-conditional measurement-tree CoMB (two memory qubits, one work qubit), rendered identically.}
		\label{fig:samples}
	\end{figure*}

\begin{figure}[b]
		\centering
		\figfile{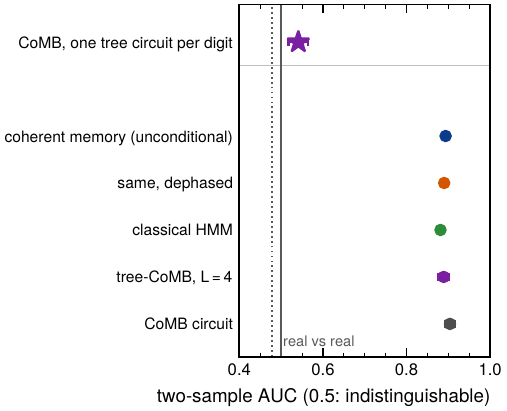}{\columnwidth}
		\caption{Two-sample AUC of generated against real digits in the stroke representation ($0.5$: indistinguishable). Star: class-conditional CoMB with one measurement-tree circuit per digit; points: unconditional models, with the standard deviation over seeds where available; dotted line: real against real digits. Error bars: statistical errors of the AUC~\cite{HanleyMcNeil1982}.}
		\label{fig:realism}
	\end{figure}

\begin{figure*}[tb]
		\centering
		\subfloat[]{\figfile{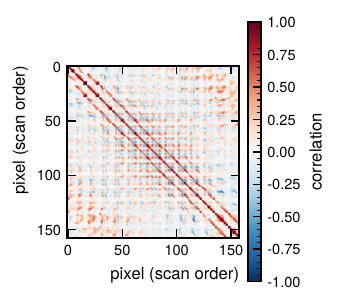}{0.33\textwidth}}
		\subfloat[]{\figfile{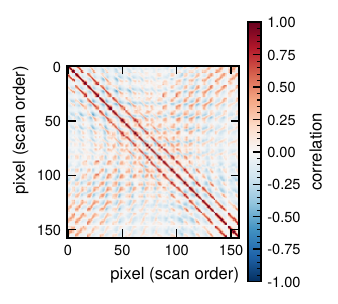}{0.33\textwidth}}
		\subfloat[]{\figfile{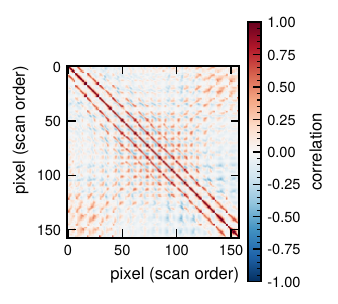}{0.33\textwidth}}
		\caption{Pixel correlation matrices of rendered digits (pixels averaged to $14\times14$, restricted to pixels that vary across test digits, in scan order): (a) test digits rendered from their strokes, (b) general coherent memory, (c) class-conditional CoMB.}
		\label{fig:pixcorr}
	\end{figure*}

	\paragraph{Sample quality.} Class-conditional generation uses one measurement-tree CoMB per digit with two memory qubits, one work qubit and node depth~4, three qubits in total. The unconditional generator is the exact mixture over classes, itself a CoMB with a classical label register (Fig.~\ref{fig:samples}). A classifier trained on real digits identifies the intended digit in $\UMixIntended$ of samples, against $\UCeil$ for real digits, and $\UDigitsPass$ of $10$ classes meet the criterion that samples be recognized at least $0.8$ times as often as real digits of that class. The two-sample AUC~\cite{HanleyMcNeil1982} of generated against real digits in the stroke representation is $\UMixAUC$, against $\UAUCfloor$ for training against test digits, so that the generated digits are close to indistinguishable from real ones in this representation (Figs.~\ref{fig:realism} and~\ref{fig:pixcorr}).
	
\begin{figure}[b]
		\centering
		\figfile{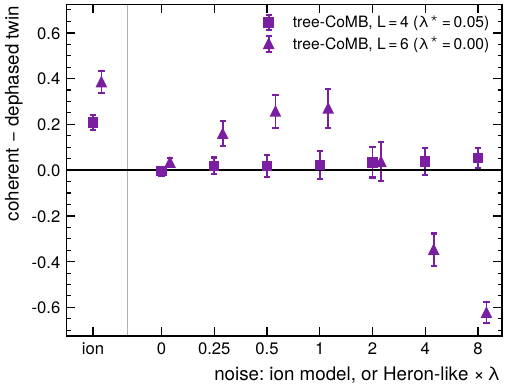}{\columnwidth}
		\caption{Trained circuits under device noise, evaluated exactly: held-out objective of each coherent circuit minus that of its dephased twin under the trapped-ion (H2-like) model, left of the gray line, and under the superconducting (Heron r3-like) model scaled by $\lambda$. Negative values indicate a coherent advantage; $\lambda^\star$ marks where the coherent tree ceases to outperform its twin. Error bars: paired jackknife errors.}
		\label{fig:noise}
	\end{figure}

	\paragraph{Device noise.} Exact evaluation under the device models of Sec.~\ref{sec:noise} shows that the coherent tree ceases to outperform its dephased twin at $\lambda^\star=\ULambdaStar$ times the nominal error rates of the superconducting model (Fig.~\ref{fig:noise}).
	
	\section{Related work}\label{sec:prior}
	
	\paragraph{Quantum generative models.} Quantum circuit Born machines~\cite{Benedetti2019,Liu2018,Coyle2020} encode a sample in the measurement of a full register and are typically trained with sample-based kernel losses~\cite{Liu2018,Rudolph2023}. CoMB retains Born-rule sampling but distributes a sample over sequential rounds of a fixed register and replaces sample-based estimates with exact losses. Classical training with quantum deployment~\cite{Recio2025} and denoising-diffusion models~\cite{Zhang2024QuDDPM} pursue scalability through different model classes. The surrogation results of Ref.~\cite{BornUltimatum2025} constrain which losses can reveal a quantum advantage. Our loss constrains the full distribution, and our controls attribute the observed advantage to coherence. Mid-circuit measurements can concentrate the output of a generator~\cite{Overconcentration2026}. The trained CoMBs emit all tokens in nearly every round and occupy their full memory, which we attribute to the round-dependent unitaries but do not prove. Systematic benchmarks find that classical baselines often outperform quantum models, and that removing entanglement often does not hurt their performance~\cite{Bowles2024}. The dephased twin is the corresponding test for CoMB, and CoMB passes it. Quantum latent distributions have improved deep generative models of molecular data~\cite{Bacarreza2025}.
	
	\paragraph{Sequential quantum models.} CoMB is a hidden quantum Markov model~\cite{Monras2010,Srinivasan2018,Adhikary2020} and a sequential generator in the sense of Refs.~\cite{Schon2005,FossFeig2021}. Its output distribution is a locally purified matrix product state of bond dimension $\chi$~\cite{Glasser2019}, which implies classical simulability at cost polynomial in $\chi$. Previous learning methods for HQMMs maximize likelihoods of discrete sequences~\cite{Srinivasan2018,Adhikary2020}. The tilted-instrument identity, which generalizes the counting-statistics generating function~\cite{levitov1996,Esposito2009,landi2024}, enables exact distribution-level training for continuous outputs. The equivalent classical memory measured here is the empirical counterpart of the quantum memory reductions established for specific processes~\cite{Gu2012,QDimRed2026}.
	
	\paragraph{Quantum memory advantages in stochastic modeling.} Quantum models of stochastic processes can store the relevant past in fewer dimensions than any classical model of equal accuracy, and for some families of processes the classical memory grows without bound while a single qubit suffices~\cite{Elliott2020}. Lossy dimension reduction extends this to approximate sampling~\cite{Yang2025}, and a single-qubit photonic memory has outperformed every classical model of the same dimension~\cite{Wu2023}. For learned models, quantum models of constrained memory are more accurate than provably optimal classical ones on discrete time series, also on quantum hardware~\cite{Yang2021}, and more accurate than classical models trained by expectation maximization on a renewal process and a short real dataset~\cite{Banchi2023}, a study that proposed testing the advantage on hardware with mid-circuit measurements. Contextuality yields provable memory separations between quantum and classical sequence models~\cite{Gao2022,Anschuetz2023} and a heuristic that predicts the performance gap of learned models, including on genomic sequences~\cite{Teo2025}. CoMB is the generative, learned counterpart for high-dimensional continuous data. Its training is exact and does not scale with the register. Quantum and classical memories are trained with the same loss and optimizer at several budgets up to convergence, the controls attribute the advantage to coherence, and the equivalent classical memory measures the dimension reduction directly. Unlike in Ref.~\cite{Yang2021}, its classical baselines are trained rather than provably optimal.
	
	\paragraph{Applications to physics data.} Existing quantum generators for calorimeter showers operate on compressed or patched representations~\cite{qfan,DualPQC2021,CaloQVAE2024}. CoMB generates the full resolution of the data set from a fixed register, and the measurement-tree compilation builds on binary-tree implementations of generalized measurements~\cite{Andersson2008} realized with dynamic circuits~\cite{Ivashkov2024}.
	
	\section{Discussion}\label{sec:discussion}
	
	The results establish three facts about the algorithm. CoMB generates high-dimensional samples, such as complete $368$-cell calorimeter showers, from a fixed register of five qubits in a single circuit execution. Such sequential generators can be trained exactly, at a cost linear in the length of the sample. Finally, the coherence of their memory measurably improves the generated distribution. On handwritten digits this improvement separates every coherent run from every dephased and every classical run, and three-qubit class-conditional models generate digits close to indistinguishable from real ones in the stroke representation. Coherence lowers the held-out loss by $23$ to $31\%$ relative to the dephased twin, at every training budget. With four memory levels, CoMB outperforms the most general classical memory of the same size at every budget, by $15\%$ at convergence. Four coherent levels do the work of about six classical states, a memory-dimension advantage seen on showers at every training budget and on handwritten digits, where every coherent run outperforms every control. With eight levels, coherent and classical memories become indistinguishable with longer training, as all models approach the floor set by the shared readout. The protocol that yields these numbers, a dephased twin, a maximal classical baseline, many seeds and a held-out loss that constrains the full distribution, is applicable to any quantum generative model.
	
	Several limitations qualify these conclusions. All models are classically simulable at a cost linear in $d$ and polynomial in $\chi$, and no computational advantage is claimed. The classical memories are trained rather than provably optimal, so the memory-dimension advantage is empirical, unlike the provable advantage for discrete time series of Ref.~\cite{Yang2021}. The generated showers do not reach the fidelity of classical fast-simulation models, so no practical quantum advantage in the sense of Ref.~\cite{HibatAllah2024} is claimed. The coherent advantage is confined to the joint distribution and is not visible in summary statistics. At $\chi=8$ the circuit's margin at the fixed budget was obtained in the third of three configurations, all of which are reported, and the two memories become indistinguishable with longer training. The digits comparisons used a fixed budget of $1\,500$ steps, at which the classical models were still improving slowly. Training to convergence ended when a block of $3\,000$ steps improved the validation loss by less than $1\%$, or after four additional blocks. The hardware experiment covers twelve of $369$ rounds and used a model trained before the directional components were introduced. Finally, the readout of eight levels per coordinate imposes a floor that no memory can lower.
	
	These limitations define the next steps. Lowering the common floor with finer readouts is the most direct test of whether the advantage reappears at larger memories. Data constructed with a controlled amount of dependence that requires positive semidefinite rather than nonnegative factorizations would test whether the advantage grows accordingly. Noise-aware training with the exact noisy loss of Sec.~\ref{sec:noise} would prepare longer executions on hardware.
	
	\section*{Code and data availability}
	
	The code that implements the model, the losses, the training, the evaluation and the device execution, and that produces every figure and table, is available at \url{https://github.com/jamalslim/comb}. The scripts for the seed studies are in the folder \texttt{maxwell/}. CaloChallenge Dataset~1~\cite{ds1} and the UCI digits~\cite{UCIdigits} are publicly available. Trained parameters and the device record (job \texttt{db34tb7r11fs7397r770}) are released with the code.
	
	\begin{acknowledgments}
		This research was supported in part through the Maxwell computational resources operated at Deutsches Elektronen-Synchrotron DESY (Hamburg, Germany), a member of the Helmholtz Association HGF. The authors acknowledge the support with funds from the Helmholtz Association HGF (Germany), Hamburgische Investitions- und F\"orderbank (IFB) (Germany), the European Union's HORIZON MSCA Doctoral Networks program project ENGAGE (101034267), and the Ministry of Science, Research and Culture of the State of Brandenburg within the Center for Quantum Technologies and Applications (CQTA) (Germany). We acknowledge the use of IBM Quantum services for this work. 
	\end{acknowledgments}

	\appendix
	
	\section{Proofs}\label{app:proofs}
	
	\paragraph{Proposition~\ref{prop:instrument}.} Measuring the work register after $U_\beta$ and discarding it maps the memory state $\rho$ to $M_{\beta,c}\rho M_{\beta,c}^\dagger$ with probability $\Tr(M_{\beta,c}\rho M_{\beta,c}^\dagger)$; iterating over rounds and multiplying the conditional probabilities gives the trace formula. Since $U_\beta$ is unitary, $\sum_cM_{\beta,c}^\dagger M_{\beta,c}=(\Id\otimes\bra{0})U_\beta^\dagger U_\beta(\Id\otimes\ket{0})=\Id$, so $\sum_c\Lam_{\beta,c}$ preserves the trace.
	
	\paragraph{Theorem~\ref{thm:tilted}.} By Proposition~\ref{prop:instrument}, $\E[\prod_tw_t(c_t)]=\sum_{c_1,\dots,c_d}\prod_tw_t(c_t)\Tr[\Lam_{d,c_d}\cdots\Lam_{1,c_1}(\rho_0)]$. The trace and the maps are linear, so each sum over $c_t$ can be carried out inside round $t$, which gives the composition of the tilted maps.
	
	\paragraph{Proposition~\ref{prop:cf}.} Given the tokens, the coordinates are independent and $z_\beta$ is uniform on bin $c_\beta$ of round $\beta$, so $\E[e^{i\omega\cdot z}\mid c]=\prod_\beta w_{\beta,c_\beta}(\omega_\beta)$ and Theorem~\ref{thm:tilted} applies. For $y=Az+b$, $\E\,e^{i\omega\cdot y}=e^{i\omega\cdot b}\,\E\,e^{i(A^T\omega)\cdot z}$.
	
	\paragraph{Corollary~\ref{cor:mmd}.} Substituting the Fourier representation of $k$ into the three expectations of the squared MMD and exchanging integrals gives $\int|\Phi_P-\Phi_Q|^2d\mu$. With frequencies drawn independently from the normalized measure, the sample average of $|\hat\Phi_N(\omega_r)-\Phi_\theta(\omega_r)|^2$ is unbiased for $\MMD_k^2(\hat P_N,Q_\theta)$, and since $\Phi_\theta$ is exact its gradient is unbiased as well; the $O(1/N)$ bias of the empirical characteristic function is removed in the final comparisons by the unbiased two-sample form.
	
	\paragraph{Proposition~\ref{prop:copula}.} The variables $t_\beta=c_\beta+u_\beta$ are continuous, with a joint distribution that is uniform on each unit cell of the token lattice, so their copula is the checkerboard copula of $c$~\cite{genest2007}; continuous strictly increasing maps leave a copula unchanged~\cite{nelsen2006}. The marginal formula follows from the uniform distribution within each cell, and inverting it at the cumulative token probabilities gives the quantile readout.
	
	\paragraph{Proposition~\ref{prop:rank}.} (i) The initial state is diagonal, and dephasing maps a diagonal state $\sum_jp_j\ket{j}\bra{j}$ after round $\beta$ and token $x$ to a diagonal state with entries $\sum_jp_j|\bra{i'}M_{\beta,x}\ket{j}|^2$, which are nonnegative and, summed over $x$ and $i'$, equal $p_j$. (ii) With $B_p=\Lam_{k,c_k}\cdots\Lam_{1,c_1}(\rho_0)$ and $E_f=(\Lam_{d,c_d}\cdots\Lam_{k+1,c_{k+1}})^\dagger(\Id)$, both positive semidefinite, $P_{p,f}=\Tr(E_fB_p)$. In the dephased model both factors are diagonal, which gives a nonnegative factorization of inner dimension $\chi$.
	
	\paragraph{Proposition~\ref{prop:contain}.} A classical HMM with transition--emission probabilities $T(k,c\mid j)$ is reproduced by the Kraus operators $M_{c,(k,j)}=\sqrt{T(k,c\mid j)}\,\ket{k}\bra{j}$, grouped by token $c$. For each token there are at most $\chi^2$ operators, so an environment of dimension $E\ge\chi^2$ suffices to dilate them to an isometry, and the operators map basis states to basis states, so no coherence is ever created.
	
	\paragraph{Theorem~\ref{thm:tree}.} Along the branch $o=b_1\cdots b_n$ the product telescopes, $N_{p_{n-1},b_n}\cdots N_{(),b_1}=V_oE_o^{1/2}E_{()}^{-1/2}=V_o|M_o|=M_o$, because $E_{()}=\Id$ and $V_o|M_o|$ is the polar decomposition of $M_o$. At every node $\sum_bN_{p,b}^\dagger N_{p,b}=E_p^{-1/2}(E_{p0}+E_{p1})E_p^{-1/2}=\Id$ on the support of $E_p$.
	
	\paragraph{Proposition~\ref{prop:nest}.} Let the added memory qubit carry zero rotation angles and take part only in controlled-$Z$ gates. On the subspace where it is in $\ket{0}$ every such gate acts as the identity. With the nested line $m_0$--$w_0$--$m_1$--$w_1$--$w_2$--$m_2$, removing $m_2$ leaves exactly the line of the smaller register, so the larger round unitary restricted to that subspace equals the smaller one; since $m_2$ starts in $\ket{0}$ and nothing moves it, the two models generate the same distribution. With the alternating line $m_0$--$w_0$--$m_1$--$w_1$--$m_2$--$w_2$, removing $m_2$ disconnects $w_2$, so no choice of the larger model's parameters reproduces the smaller one.
	
	\section{Implementation and checks}\label{app:checks}
	
	Every quantity was checked against an independent reference before it was used. For the token models, the window and pair laws agree with a brute-force enumeration of all token sequences for the circuit, its twin and the HMM to $6\times10^{-16}$; the path characteristic function agrees with an enumeration of $16^4$ sequences to $4\times10^{-14}$; every gradient agrees with central finite differences to $10^{-8}$; the batched node circuits agree exactly with a single-circuit simulation, and the noisy tree with a node-by-node density-matrix simulation to $6\times10^{-16}$; Theorem~\ref{thm:tree} holds on random general memories to $1.4\times10^{-15}$; and the samplers agree with the exact laws, within statistical noise, on $6\times10^4$ samples. For the shower models, the round instruments are completely positive and trace preserving to $10^{-15}$; the exact characteristic function agrees with the empirical one of $20\,000$ sampled showers within the statistical scale; analytic and finite-difference gradients agree to three digits, including on the angles of the qubits the nested wiring moves; and the nested and the interleaved eight-level circuits contain or fail to contain the trained four-level model as described in Appendix~\ref{app:proofs}, to $3\times10^{-17}$ and $7\times10^{-4}$ respectively. After every change to the code we verified that the default model trains bit for bit as before. The device circuit was compared with the model's instruments to $10^{-15}$ before submission.
	
	\section{Protocols}\label{app:protocols}
	
	\paragraph{Data.} The UCI digits are split $1\,200/297/300$ into training, validation and test sets with seed $2026$. The sequence length is the $90$th percentile of the walk lengths, rounded up to a multiple of $8$ and capped at $128$. The shower data are split as in Sec.~\ref{sec:ds1}.
	
	\paragraph{Statistical errors.} On the digits every held-out result compares a model with a finite set of test digits, and its error is that of a delete-a-group jackknife: the test digits are split into $50$ groups and the result is recomputed with each left out in turn. All models are scored on the same digits, so differences carry paired errors, much smaller than the common uncertainty of the absolute values, $\pm\UCommonSE$ for the coherent memory. On the showers the errors are standard errors over $30$ training seeds; all models are scored on the same $2\,000$ held-out showers.
	
	\paragraph{Training.} On the digits we use Adam for $1\,500$ steps with $50$ warm-up steps and a cosine schedule, with learning rates of $0.1$ for circuits, $0.05$ for general memories and $0.2$ for HMMs, and select the step with the lowest exact validation objective, evaluated every $25$ steps. Path terms use $256$ fresh frequencies per step from a Gaussian mixture with bandwidths $\sigma/2$, $\sigma$ and $2\sigma$, $\sigma$ the median distance. On the showers we use Adam for $3\,000$ steps with peak learning rates $0.1$ for the angles and $0.03$ for the readouts, $50$ warm-up steps and cosine decay, and $R=256$ frequencies per step from a pool of $32\,768$ orthogonal random features of the Gaussian mixture, extended by $8\,192$ frequencies along the energy fractions of the two dense layers with magnitudes $r/\sigma_a$, $r\in\{2,4\}$; validation every $25$ steps on $2\,000$ showers, best step kept. For the $6\,000$-step budget, each trained model is continued from its parameters for $3\,000$ further steps with $0.1$ times the peak learning rates and the same schedule shape, and the best step, including the starting point, is kept. The general coherent memory is started from each trained circuit by writing every round unitary as $e^{-iH}$ with $H=i\log U$, which reproduces the circuit's characteristic function to $10^{-17}$. Its learning rates are those of the circuits divided by $\sqrt{D}$, with $D$ the dimension of the register. For training to convergence, models are continued in the same way in blocks of $3\,000$ steps until a block improves the best validation loss by less than $1\%$, with at most four additional blocks. Seeds: $30$ each for CoMB, its dephased twin and the HMM with $\chi=4$; $30$ each for CoMB and its twin with $\chi=8$ with the alternating wiring and five layers, with the nested wiring and five layers, and with the nested wiring and $18$ layers, and for the HMM with $\chi=8$; $10$ each for HMMs with $\chi=6$, $12$, $16$, $24$, $32$, $48$ and $64$. The continued and converged comparisons use $30$ seeds each for the four-level circuit and its twin, for the general coherent memory and its twin at $\chi=4$ and $\chi=8$, and for the HMMs with $\chi=4$ and $8$, and $10$ seeds for the HMMs with $6$ to $32$ states, together with $48$ and $64$ states for the continued HMMs. Intervals of the equivalent classical memory resample the seeds of the coherent model and of every classical size with replacement, $10\,000$ times, and repeat the interpolation of Definition~\ref{def:equiv}. The held-out MMD$^2$ uses $4\,096$ frequencies on $2\,000$ test showers. The shower runs were made on the DESY Maxwell cluster.
	
	\paragraph{Noise models.} The trapped-ion, H2-like model has $p_2=10^{-3}$, $p_1=2\times10^{-5}$, readout and reset errors of $10^{-3}$, and memory dephasing of $1.4\times10^{-3}$ per window. The superconducting, Heron r3-like model has $p_2=1.5\times10^{-3}$, $p_1=2.5\times10^{-4}$, readout errors of $1.3\times10^{-2}$, reset errors of $10^{-2}$, $T_1=T_2=200~\mu$s, and $2~\mu$s per measure-and-reset window. The values follow published device characterisations, except that $p_1$, the reset error and the window duration of the superconducting model are assumptions. Each node window of a tree is charged the same idle time as a full round of the monolithic circuit.
	
	\section{Numerical results}\label{app:numbers}
	
	Table~\ref{tab:results} gives the numerical values plotted in Figs.~\ref{fig:memory} and~\ref{fig:circuits}. The full objective is the sum of its local part, the $k$-point MMD of windows and pairs of moves, and its global part, the path MMD of the pen position. For the memories the table gives the mean and standard deviation over three seeds. This spread reflects training, not the test set, which is why the main text compares models through paired differences on the same test digits, whose jackknife errors are much smaller than the spread of the absolute values.
	
	Two patterns stand out. First, the coherent memory is not only better on average but far more consistent. Its spread over seeds is $0.002$, against $0.018$ for the dephased memory and $0.037$ for the classical HMM. Second, its advantage lies in the local part of the objective, $0.577$ against $0.684$ and $0.710$, whereas the classical HMM matches the global pen path slightly better, $0.180$ against $0.209$. The circuits show the same division. Each beats its dephased twin on the local part and loses on the global part, which is why the full objectives of the trees and their twins differ little. The memory without environment levels reaches $0.788$, the same as with them, so circuits of sufficient depth would not need environment qubits for these data. The last column shows that every unconditional model is easily told apart from real digits, with AUC between $0.88$ and $0.91$. The class-conditional mixture of Sec.~\ref{sec:digits} reduces this to $0.541$.
	
	Table~\ref{tab:hmmcurves} lists the classical curves from which every equivalent classical memory for the showers is read, the held-out loss of the classical HMMs at each number of states and each training budget. At $3\,000$ steps the curve flattens at $1.35$ to $1.40\times10^{-4}$ from $24$ states on, a limit of the budget. At convergence it flattens at $1.08$ to $1.10\times10^{-4}$ from $12$ states on, the floor of the readout. The values are given to three decimals, so that the equivalent memories and their intervals in Sec.~\ref{sec:matched} can be recomputed exactly.
	
	\begin{table}[t]
		\centering\small\setlength{\tabcolsep}{4pt}
		\begin{tabular}{rcccc}
			\toprule
			$\chi$ & seeds & $3\,000$ steps & $6\,000$ steps & converged\\
			\midrule
			$4$ & $30$ & $2.805\pm0.086$ & $1.986\pm0.053$ & $1.476\pm0.028$\\
			$6$ & $10$ & $2.082\pm0.069$ & $1.511\pm0.039$ & $1.247\pm0.038$\\
			$8$ & $30$ & $1.860\pm0.024$ & $1.372\pm0.015$ & $1.162\pm0.015$\\
			$12$ & $10$ & $1.596\pm0.051$ & $1.225\pm0.017$ & $1.083\pm0.012$\\
			$16$ & $10$ & $1.536\pm0.040$ & $1.243\pm0.037$ & $1.087\pm0.018$\\
			$24$ & $10$ & $1.404\pm0.013$ & $1.195\pm0.013$ & $1.097\pm0.016$\\
			$32$ & $10$ & $1.384\pm0.017$ & $1.194\pm0.016$ & $1.079\pm0.012$\\
			$48$ & $10$ & $1.346\pm0.014$ & $1.186\pm0.013$ & --\\
			$64$ & $10$ & $1.362\pm0.021$ & $1.179\pm0.019$ & --\\
			\bottomrule
		\end{tabular}
		\caption{Held-out MMD$^2$ ($10^{-4}$) of the classical HMMs on the showers, mean $\pm$ standard error over seeds, at the three training budgets of Appendix~\ref{app:protocols}. These curves define the equivalent classical memory at each budget (Definition~\ref{def:equiv}). The converged curve was computed up to $32$ states and is flat from $12$ states on.}
		\label{tab:hmmcurves}
	\end{table}
	
	\begin{table}[t]
		\centering\small\setlength{\tabcolsep}{3pt}
		\fitcol{\begin{tabular}{lcccc}
				\toprule
				model & held-out & local & global & AUC\\
				\midrule
				\multicolumn{5}{l}{\emph{memories, $\chi=4$}}\\
				coherent memory, 16 env.\ levels & $0.786\pm0.002$ & $0.577\pm0.002$ & $0.209\pm0.001$ & 0.893\\
				same memory, dephased & $0.890\pm0.018$ & $0.684\pm0.020$ & $0.206\pm0.002$ & 0.890\\
				classical HMM & $0.890\pm0.037$ & $0.710\pm0.043$ & $0.180\pm0.006$ & 0.881\\
				coherent memory, no env.\ levels & $0.788$ & $0.580$ & $0.208$ & 0.892\\
				\multicolumn{5}{l}{\emph{circuits of fixed depth}}\\
				CoMB circuit, 4 layers & $1.292$ & $1.098$ & $0.194$ & 0.904\\
				\quad its dephased twin & $1.360$ & $1.176$ & $0.184$ & 0.907\\
				tree-CoMB, node depth 4 & $0.815$ & $0.614$ & $0.201$ & 0.889\\
				\quad its dephased twin & $0.820$ & $0.675$ & $0.145$ & 0.895\\
				tree-CoMB, node depth 6 & $0.802$ & $0.595$ & $0.207$ & 0.884\\
				\quad its dephased twin & $0.769$ & $0.626$ & $0.142$ & 0.879\\
				\bottomrule
		\end{tabular}}
		\caption{Held-out results on the test digits (lower is better): full objective, its local ($k$-point) and global (path) parts, and two-sample AUC of rendered samples against real digits ($0.5$: indistinguishable). Memories: mean $\pm$ standard deviation over three seeds (one seed for the memory without environment levels); circuits: seed $0$. Figures~\ref{fig:memory} and~\ref{fig:circuits} plot these values.}
		\label{tab:results}
	\end{table}

\end{document}